\documentclass[a4paper,11pt]{article}
\pdfoutput=1
\usepackage[utf8]{inputenc}
\usepackage{jcappub}
\usepackage[T1]{fontenc}

\usepackage[mathscr]{eucal}
\usepackage[Symbolsmallscale]{upgreek}

\newcommand{\Thc}{T_\text{hc}}

\newcommand{\OGW}{\Omega_\text{GW}}

\title{Primordial Gravitational Wave Signals in Modified Cosmologies}

\author[a]{Nicolás Bernal,}
\author[b]{Anish Ghoshal,}
\author[c]{\\Fazlollah Hajkarim}
\author[d,\,e]{and Gaetano Lambiase}

\affiliation[a]{Centro de Investigaciones, Universidad Antonio Nariño,\\
Carrera 3 Este \# 47A-15, Bogotá, Colombia}
\affiliation[b]{ I. N. F. N. -  Rome Tor Vergata,
via della Ricerca Scientifica, I-00133 Rome, Italy}
\affiliation[c]{Institut für Theoretische Physik, Goethe Universität,\\
Max-von-Laue-Straße 1, D-60438, Frankfurt, Germany}
\affiliation[d]{Dipartimento di Fisica ``E.R. Caianiello'', Università di Salerno, I-84084 Fisciano (SA), Italy}
\affiliation[e]{INFN - Sezione di Napoli, Gruppo collegato di Salerno, I-84084 Fisciano (SA), Italy}

\emailAdd{nicolas.bernal@uan.edu.co}
\emailAdd{aghoshal@roma2.infn.it}
\emailAdd{hajkarim@th.physik.uni-frankfurt.de}
\emailAdd{lambiase@sa.infn.it}

\abstract{
Modified expansion rates in the early Universe prior to big bang nucleosynthesis are common in modified gravity theories, and can have a significant impact on the generation of dark matter, matter-antimatter asymmetry, primordial black holes and the primordial gravitational wave (PGW) spectrum.
Here we study the PGW spectrum in modified gravity theories, in early Universe cosmology. In particular, we consider scalar-tensor and extradimensional scenarios, investigating the detection prospects in current and future GW observatories. 
For the scalar-tensor case, PGW could be potentially observed by laser interferometers operating in the high-frequency range,  while for the extradimensional case they could be detected even at low frequencies with pulsar timing arrays. We find that data from the planned network of several GW detectors operating across various frequency ranges could be able to distinguish between various modified gravity scenarios.
}

\begin{document}

\begin{flushright}
    PI/UAN-2020-677FT
\end{flushright}

\maketitle


\section{Introduction} 
\label{inro}

Probes of the early Universe include measures of the primordial abundances of light elements generated during big bang nucleosynthesis (BBN), cosmic microwave background (CMB) radiation, etc. Recently, the  detection of gravitational waves (GW) by LIGO~\cite{TheLIGOScientific:2014jea} and Virgo~\cite{TheVirgo:2014hva} has opened up the opportunity to probe fundamental physics, otherwise not accessible via other interactions. Hitherto all the GW signals observed so far  are of astrophysical origin, e.g. compact binary systems~\cite{Abbott:2016blz, LIGOScientific:2018mvr}, but in a foreseeable future we expect to be able to explore the same of cosmological origin. In particular, primordial GW (PGW) could originate  from quantum fluctuations during the inflationary period of the early Universe or phase transitions~\cite{Maggiore:1999vm}. 

The propagation of PGW carries information of the expansion history of the Universe through its evolution in the post-inflationary phase. Thus, such GW serve as a useful tool to probe the cosmological history of our Universe prior to BBN, for example, the reheating  temperature~\cite{Gouttenoire:2019kij, Auclair:2019wcv}, the equation-of-state parameter~\cite{GarciaBellido:2007af}, the  quark-hadron phase  transition in  QCD~\cite{Hajkarim:2019csy, Hajkarim:2019nbx}, and properties of possible hidden sectors beyond the standard model (SM)~\cite{Giudice:2016zpa}.

We investigate this probe of the early Universe expansion rate to understand PGW signals predicted in various modified gravity theories of the early Universe. Modifications of general relativity (GR) naturally imply variations to the Hubble expansion rate of the Universe from the metric level of the theory, leaving an imprint on any processes that carry information of the expansion history, and have potential observational signatures. 
However, the success of BBN strongly favors a standard history for temperatures below $\sim 4$~MeV~\cite{Kawasaki:2000en, Hannestad:2004px, deSalas:2015glj}.
Nonstandard cosmological scenarios can affect the dark matter (DM) relic density~\cite{Catena:2004ba, Catena:2009tm, Meehan:2015cna, Dutta:2016htz, DEramo:2017gpl, DEramo:2017ecx, Bernal:2018ins, Lambiase:2018yql, Bernal:2018kcw, Arias:2019uol, Poulin:2019omz, Bernal:2019mhf}, the matter-antimatter asymmetry~\cite{Kamada:2019ewe,Okada:2005kv, Lambiase:2006dq, Barenboim:2012nh, Lambiase:2013haa, deCesare:2014dga, Bernal:2017zvx, Dutta:2018zkg}, the spectrum of PGW~\cite{Bettoni:2018pbl, DEramo:2019tit, Bernal:2019lpc, Figueroa:2019paj, Bhattacharya:2019bvk, Gouttenoire:2019kij,Guo:2020grp} and the abundance of primordial black holes~\cite{Khlopov:1980mg, Polnarev:1986bi, Georg:2016yxa, Harada:2016mhb, Cotner:2016cvr, Harada:2017fjm, Kokubu:2018fxy} and  microhalos~\cite{Erickcek:2011us, Barenboim:2013gya, Fan:2014zua, Delos:2018ueo, Redmond:2018xty}.
See Ref.~\cite{Allahverdi:2020bys} for a recent review.

The fact that Einstein's theory of gravity breaks down in the UV motivates the consideration of modifications of GR~\cite{Capozziello:2011et, Will:2018bme}.
Deviations from GR follow in different frameworks, such as Brans-Dicke and scalar-tensor (ST) theories~\cite{Perivolaropoulos:2009ak, Hohmann:2013rba, Jarv:2014hma}, braneworld theories~\cite{Nojiri:2002wn, Bronnikov:2006jy, Kaminski:2009dh, Benichou:2011dx, Guo:2014bxa, Donini:2016kgu}, $f(R)$ and $f(\phi,\,R,\,R^2)$ theories~\cite{Berry:2011pb, Capozziello:2014mea, Lambiase:2015yia, Schellstede:2016ldu}, noncommutative geometry~\cite{Lambiase:2013dai}, and compactified extra dimension/Kaluza-Klein models~\cite{ArkaniHamed:1998rs, ArkaniHamed:1998nn, Antoniadis:1998ig, Floratos:1999bv, Kehagias:1999my, Perivolaropoulos:2002pn}.
Moreover, they can also be generated from higher-order terms in the curvature invariants, nonminimal couplings to the background geometry in the Hilbert-Einstein Lagrangian~\cite{Birrell:1982ix, Gasperini:1991ak, Vilkovisky:1992pb, Nojiri:2006ri} or to curvature invariants such as $R^{2}$, $R_{\mu\nu} R^{\mu\nu}$, $R^{\mu\nu\alpha\beta}R_{\mu\nu\alpha\beta}$, $R \,\Box R$, or $R\,\Box^{k}R$, corresponding equivalently to Einstein's gravity plus one or multiple conformally coupled scalar fields~\cite{Teyssandier:1983zz, Maeda:1988ab, Wands:1993uu, Schmidt:2001ac}. 
Additional terms into the action of gravity may also come from string loop effects~\cite{Damour:1994ya}, dilaton fields in string cosmology~\cite{Gasperini:1994xg}, and nonlocally  modified  gravity induced by  quantum  loop  corrections~\cite{Deser:2007jk}.

In this article, we study PGW spectra in modified gravity theories, namely, ST gravity and braneworld cosmology. We identify parameter spaces that give a substantial boost to the PGW spectrum to be detectable by current and future GW detectors, thereby providing a test for
deviations from GR in the early Universe. 

The paper is arranged as follows. In Sec.~\ref{secstanpgw} we describe the generation and evolution of PGW in the standard Friedmann-Lemaître-Robertson-Walker (FLRW) cosmology.
In Sec.~\ref{sec:modgragen}, a general parametrization for different theories of gravity affecting the expansion of the Universe is described. 
Next, we present estimations of PGW in ST (Sec.~\ref{sec:sttpgw}) and braneworld (Sec.~\ref{branegw}) cosmological scenarios to understand the propagation of GW in early Universe.
Finally, we conclude in Sec.~\ref{sec:conclusion}.

\section{PGW in standard cosmology}
\label{secstanpgw}
In this section, we briefly review the computation of the PGW spectrum in the standard cosmological scenario.
GWs correspond to spatial metric perturbations satisfying the transverse traceless conditions: $\partial^ih_{ij}=0$ and $h_i^i=0$.
The equation of motion for tensor perturbation at the first order of cosmic perturbation theory can be written as~(following Refs.~\cite{Watanabe:2006qe, Saikawa:2018rcs, Bernal:2019lpc})
\begin{equation}\label{eq:gw-pert}
    \ddot h_{ij}+3H\,\dot h_{ij}-\frac{\nabla^2}{a^2}h_{ij}=16\pi\,G\,\Pi_{ij}^\text{TT}\,,
\end{equation}
where the dots correspond to derivatives with respect to the cosmic time $t$, and $G$ is the Newton's constant.
The conformal time $\tau$ is related to the standard time as $dt=a\,d\tau$ and  $a^\prime=a^2 H$, where the prime corresponds to a derivative with respect to $\tau$. 
The Hubble expansion rate $H$ in GR is given by 
\begin{equation}\label{hubble}
    H_\text{GR}\equiv\frac{\dot{a}}{a}=\sqrt{\frac{8\pi}{3}G\,\rho}\,,
\end{equation}
where 
\begin{equation}
    \rho(T)\equiv \frac{\pi^2}{30}\,g(T)\,T^4
\end{equation}
is the SM energy density, and $g(T)$ corresponds to the effective number of relativistic degrees of freedom of SM radiation, as a function of the SM temperature $T$.
Here we use the data of Ref.~\cite{Drees:2015exa} to consider the effect of  the thermal evolution of SM degrees of freedom. This effect is mostly important around the QCD epoch  $T\simeq 150$~MeV and the electroweak transition $T\simeq 100$~GeV  where the Higgs, the electroweak gauge bosons and the top quark  decoupling happens.
In fact, the effects of the QCD equation of state and the lepton asymmetry on PGW can have an impact up to a few percent around the QCD and electroweak epochs~\cite{Hajkarim:2019csy}.
Finally, let us add that in the right hand side of Eq.~\eqref{eq:gw-pert}, $\Pi_{ij}^\text{TT}$ is the transverse-traceless part of the anisotropic stress tensor $\Pi_{ij}$ defined as
\begin{equation}
    \Pi_{ij}\equiv \frac{T_{ij}-p\,g_{ij}}{a^2}\,,
\end{equation}
where $T_{ij}$ is the stress-energy tensor, $g_{ij}$ the metric tensor, and $p$ the background pressure.
$\Pi_{ij}^\text{TT}$ can be the source for tensor perturbations at frequencies smaller than $10^{-10}$~Hz (corresponding to temperatures $T\lesssim 4$~MeV) due to the free streaming of neutrinos and photons~\cite{Weinberg:2003ur, Saikawa:2018rcs}.
However, here we focus on a higher frequency range and therefore this term will be disregarded. 
To solve the tensor perturbation equations, one can rewrite it in the Fourier space as \cite{Watanabe:2006qe, Saikawa:2018rcs, Bernal:2019lpc}
\begin{equation}
    h_{ij}(t,\,\vec x)=\sum_\lambda\int\frac{d^3k}{(2\pi)^3}\,h^\lambda(t,\,\vec k)\,\epsilon_{ij}^\lambda(\vec k)\,e^{i\,\vec k\cdot\vec x}\,,
\end{equation}
where $\lambda=+$, $\times$ corresponds to the two independent polarization states, and $\epsilon^\lambda$ is the spin-2 polarization tensor satisfying the normalization condition $\sum_{ij}\epsilon_{ij}^\lambda\epsilon_{ij}^{ \lambda'*}=2\delta^{\lambda\lambda'}$.
The tensor perturbation can be written in terms of 
\begin{equation}
    h^\lambda(t,\vec{k})=h^\lambda_\text{prim}(\vec{k})\,X(t,k)\,,
\end{equation}
where $X$ is a transfer function and $h^\lambda_{\text{prim}}(\vec{k})$ is the amplitude of the primordial tensor perturbations.
The tensor power spectrum can be expressed as \cite{Watanabe:2006qe, Saikawa:2018rcs, Bernal:2019lpc}
\begin{equation}
    \mathcal{P}(k)=\frac{k^3}{\pi^2}\sum_{\lambda}\left|h^\lambda_{\text{prim}}(\vec{k})\right|^2=\left.\frac{2}{\pi^2}\,G\,H^2\right|_{k=a\,H}.
\end{equation}
Equation~\eqref{eq:gw-pert} can therefore be rewritten as 
\begin{equation}\label{eq:gwode}
    X^{\prime\prime}+2\frac{a^{\prime}}{a}\,X^\prime+k^2 X=0\,,
\end{equation}
and behaves like a damped oscillator.
Let us note that, as the scale factor and the conformal time are related via the equation of state as $a\propto \tau^{\frac{2}{1+3\omega}}$, the damping term in Eq.~\eqref{eq:gwode} can be rewritten as
\begin{equation}\label{eq:friction}
    2\,\frac{a'}{a}=\frac{4}{1+3\omega}\,\frac{1}{\tau}\,.
\end{equation}

In standard cosmology, the relic density of PGW from first-order tensor perturbation becomes \cite{Watanabe:2006qe, Saikawa:2018rcs, Bernal:2019lpc}
\begin{equation}\label{pgw-relic}
    \OGW(\tau,\,k)=\frac{\mathcal{P}_T(k)\,\left[X^\prime(\tau,\,k)\right]^2}{12\,a^2(\tau)\,H^2(\tau)}
    \simeq \frac{1}{24}\mathcal{P}_T(k)\left[\frac{a_\text{hc}}{a(\tau)}\right]^4\left[\frac{H_\text{hc}}{H(\tau)}\right]^2,
\end{equation}
where in the last step we used the fact that after averaging over periods of oscillations $X^\prime(\tau,\,k)\simeq k\,X(\tau,\,k)\simeq k\, a_\text{hc}/(\sqrt{2}\,a(\tau))\simeq   a^2_\text{hc}\,H_\text{hc}/(\sqrt{2}\,a(\tau))$, with $k=2\pi\,f=a_\text{hc}\,H_\text{hc}$ for the horizon crossing moment.%
\footnote{At super-horizon scales ($k\ll a\,H$) the transfer function $X(\tau,\,k)\rightarrow 1$. The initial conditions for the transfer function (at superhorizon scale) can be written as
\begin{equation}
\label{init}
    X(0,\,k)=1\,,\qquad\qquad
    X^\prime(0,\,k)=0\,.
\end{equation}
Since the numerical computation of Eq.~\eqref{eq:gwode} from the early Universe before horizon crossing until today is numerically expensive, one can use WKB approximation. 
After horizon crossing (hc) 
\begin{equation}
    X(\tau,\,k)=\frac{\mathcal{X}}{a(\tau)}\,\sin(k\,\tau+\delta)\,,
\end{equation}
where the parameters $\mathcal{X}$ and $\delta$ can be fixed using WKB approximation for $X$ and its derivative at horizon crossing.}
The PGW relic density today ($\tau=\tau_0$) as a function of the wave number is therefore
\begin{equation}\label{gwrelic-smdof}
    \OGW(\tau_0,\,k)\,h^2\simeq \frac{1}{24}\,\left[\frac{g(\Thc)}{2}\right]\,\left[\frac{h(T_0)}{h(\Thc)}\right]^{4/3}\mathcal{P}_T(k)\,\Omega_{\gamma}(T_0)\,h^2,
\end{equation} 
where $h(T)$ corresponds to the effective number of relativistic degrees of freedom contributing to the SM entropy density $s$ defined by~\cite{Drees:2015exa}
\begin{equation}
    s(T)\equiv \frac{2\pi^2}{45}\,h(T)\,T^3\,.
\end{equation}
The scale dependency of the tensor power spectrum is given by
\begin{equation}
    \mathcal{P}_T(k)=A_T\left(\frac{k}{\tilde{k}}\right)^{n_T},\qquad A_T=r\,A_S\,,
\end{equation}
where $\tilde{k}=0.05$~Mpc$^{-1}$ is a characteristic pivot scale, and the tensor spectral index $n_T$. 
The amplitude of the tensor perturbation is denoted by $A_T$, which is written in terms of the tensor-to-scalar ratio $r$ and the scalar perturbation amplitude $A_S$. The $\texttt{PLANCK}$ mission has measured $A_S\simeq2.1\times 10^{-9}$ at the CMB scale, and put an upper bound on $r\lesssim 0.07$~\cite{Aghanim:2018eyx}.

\begin{figure}[t]
    \begin{center}
        \includegraphics[height=0.5\textwidth]{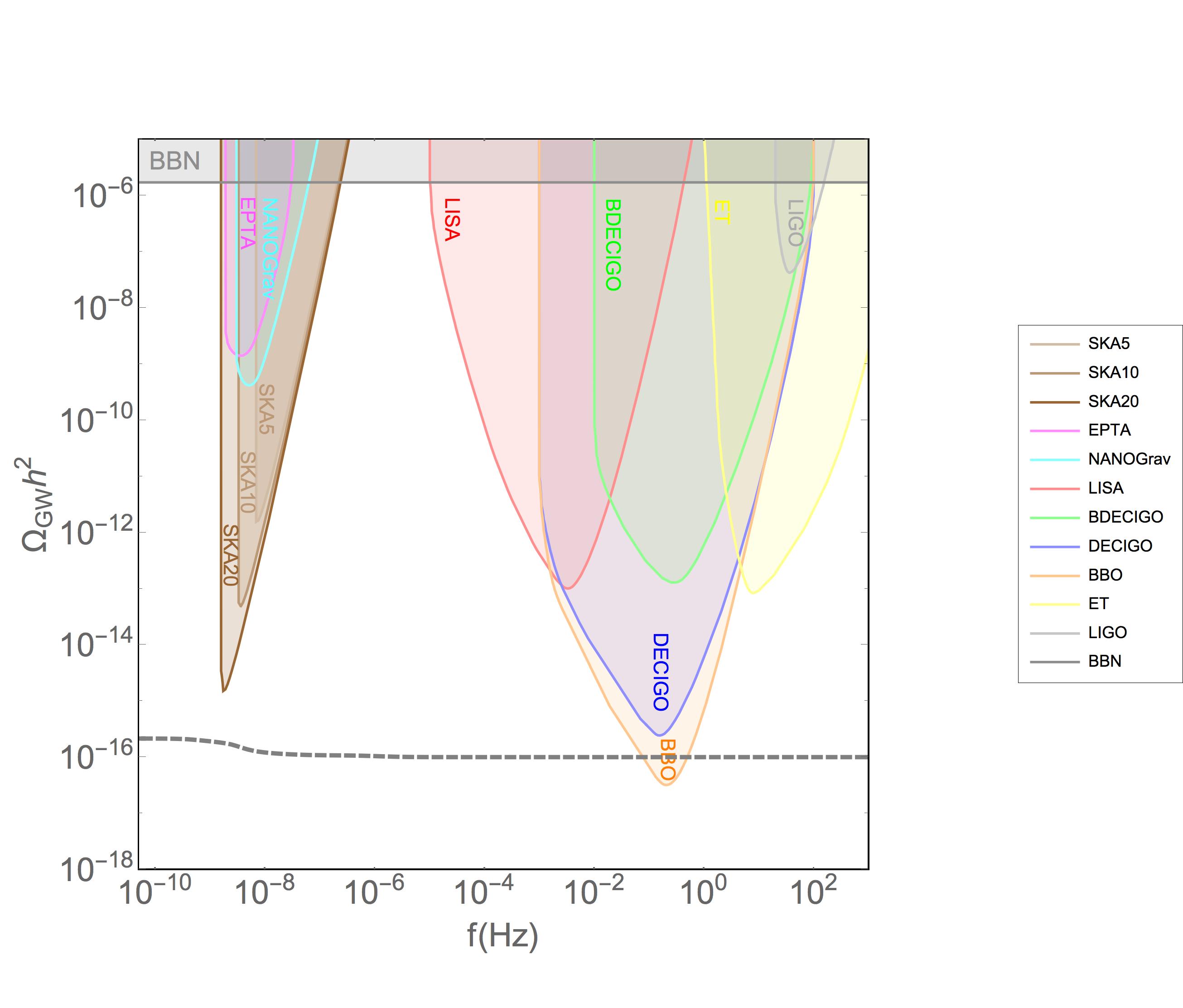}
        \caption{\it  PGW spectrum $\Omega_\text{GW}h^2$ as a function of the frequency $f$ assuming a scale invariant primordial tensor spectrum ($n_T = 0$) with a tensor-to-scalar ratio $r=0.07$, for the standard cosmological scenario.
        The colored regions correspond to projected sensitivities for various GW observatories, and to the BBN constraint described in the text.
        }
        \label{fig:pgwsm-exp}
    \end{center}
\end{figure}

Figure~\ref{fig:pgwsm-exp} represents  an example of a PGW spectrum $\Omega_\text{GW}h^2$ versus the frequency $f$ with a dashed gray line, assuming a primordial tensor spectrum with $n_T = 0$, $A_S=2.1\times 10^{-9}$ and $r=0.07$, computed following Eq.~\eqref{gwrelic-smdof}, for the standard cosmological scenario.
Additionally, the colored regions correspond to projected sensitivities for various GW observatories~\cite{Breitbach:2018ddu}.
In particular, we consider constraints from the  space-based LISA~\cite{Audley:2017drz} interferometer, the ground-based Einstein Telescope detector (ET)~\cite{Sathyaprakash:2012jk} as well as the  successor experiments BBO~\cite{Crowder:2005nr} and (B-)DECIGO \cite{Seto:2001qf, Kawamura:2020pcg}.
Moreover, we include pulsar timing arrays, in particular the currently operating  NANOGrav~\cite{Arzoumanian:2018saf} and EPTA~\cite{Lentati:2015qwp}, as well as the future SKA~\cite{Janssen:2014dka} telescope. The BBN bound comes from the constraint on the number of effective neutrinos, a variation of which affects the primordial element abundances measurements during BBN ~\cite{Boyle:2007zx, Stewart:2007fu, Kohri:2018awv}.


\section{PGW in modified cosmological histories: generalized study}
\label{sec:modgragen}

Hereafter we only consider modification of gravity due to the change of Hubble rate at subhorizon scales, which is compatible with LIGO observations of binary neutron stars and black hole mergers. In fact, the tensor perturbation equation in a modified gravity scenario may produce GW with a speed different from the speed of light~\cite{Burrage:2016myt, Monitor:2017mdv,Aoki:2020iwm,Cai:2015yza,Cai:2016ldn}. However, due to the constraint from LIGO, we study modified gravity scenarios that can mainly modify the Hubble rate~\cite{Dunsby:1998hd, Lin:2016gve, Nunes:2018zot}. Additionally, using the constraints on the tensor-to-scalar ratio and the scalar spectral index~\cite{Kinney:2005in, Adshead:2010mc} the variation of the number of $e$-folds from inflationary phase is constrained by the CMB observation to be $|\Delta N_{\text{inf}}|\lesssim 10$~\cite{Akrami:2018odb}. Any modification of GR or nonstandard cosmology in the pre-BBN era should satisfy this bound as explained in Ref.~\cite{Bernal:2019lpc}. Modified gravity dominated eras during the pre-BBN epoch that we consider here naturally respect this bound. 

We will not investigate modified gravity theories affecting the standard cosmology during the post-BBN era and especially the formation of structures at large scales. Depending on the details of a modified cosmology scenario, the primordial density perturbations might grow that can boost the formation of small  structures~\cite{Redmond:2017tja, Redmond:2018xty,Bernal:2019lpc}. Modified gravity can enhance the density perturbation at small scales, however, due to Silk damping these perturbations will dilute and not be effective at the time of structure formation which happens around $\sim 1$~eV (matter-radiation equality)~\cite{Silk:1967kq, Artymowski:2016tme}.

\subsection{Parametrization}
Modifications to GR lead to cosmological histories with expansion rates $H$ of the Universe larger
than the Hubble expansion rate $H_\text{GR}$ of standard cosmology.
The expansion rate of the Universe in modified cosmologies can be parameterized as follows~\cite{Modak:1999nm, Schelke:2006eg, Catena:2009tm, Dent:2009bv, Leon:2013qh}
 \begin{equation}\label{H=AHGRIce}
   H(T)\equiv A(T)\,H_\text{GR}(T)\,,
 \end{equation}
where $A(T)$ is the so-called amplification factor.
Since the pre-BBN epoch is not directly constrained by cosmological observations, for temperatures larger than $T_\text{BBN}$, $A(T)$ can be significantly different from unity, leading to modified cosmological expansion histories.
However, after BBN (i.e. $T\leq T_\text{BBN}$) the standard cosmology should be at work. According to this, the amplification factor $A(T) \neq 1$ at early times, and $A(T)\to 1$ at the onset of the BBN period.
Typically, it is parameterized as
 \begin{equation}\label{A(T)Ice}
   A(T)= 1 + \eta\left(\frac{T}{T_\star}\right)^\nu,
 \end{equation}
where $T_\star$ is a temperature scale, and $\eta$ and $\nu$ are dimensionless parameters, all depending on the specific cosmological model under consideration.
Alternatively, it can also be written as~\cite{Catena:2009tm}
 \begin{equation}\label{A(T)}
   A(T)=\left\{ \begin{array}{lcr}
    1+\eta\left(\frac{T}{T_\star}\right)^\nu \tanh \frac{T-T_\text{re}}{T_\text{re}} \qquad & \mbox{for} & T>T_\text{re}\,,\\
   1 & \mbox{for} & T\leq T_\text{re}\,,
   \end{array} \right.
 \end{equation}
where $T_{\star}\geq T_\text{re} > T_\text{BBN}$.
In the limit $T\gg T_\text{re}$ and $\nu\geq0$, the parametrization in Eqs.~\eqref{A(T)Ice} and~\eqref{A(T)} coincide.
However, the latter has the advantage of allowing the exploration of negative values for $\nu$.

Different values for the parameter $\nu$ appear in various modified cosmological scenarios~\cite{Modak:1999nm, Schelke:2006eg, DAmico:2009tep, Catena:2009tm, Dent:2009bv, Leon:2013qh}: $\nu=2$ in Randall-Sundrum type II brane cosmology~\cite{Randall:1999vf}, $\nu=1$ in kination models~\cite{Salati:2002md, Pallis:2005hm, Guo:2009nt}, $\nu=0$ in cosmologies with an overall boost of the Hubble expansion rate like in the case of a large number of additional relativistic 
degrees of freedom in the thermal plasma~\cite{Catena:2009tm},
$\nu=2/n-2$ in $f(x)$ cosmology with $f(x)=x+\alpha\,x^n$, where $x=R$, ${\cal T}$; $R$ and ${\cal T}$ being the scalar curvature and the scalar torsion, respectively~\cite{Capozziello:2008rq, Capozziello:2015ama, Cai:2015emx, Capozziello:2017bxm}.%
\footnote{See Refs.~\cite{Nojiri:2017ncd, Sotiriou:2008rp} for cosmological and further theoretical motivations for such theories.}

If the evolution of the Universe is adiabatic, and therefore the SM entropy is conserved, the temperature and the scale factor are related via
\begin{equation}\label{tempsca}
	\frac{dT}{da}=-\frac{1}{1+\frac{T}{3h}\frac{dh}{dT}}\frac{T}{a}\,.
\end{equation}
The nontrivial behavior due to the variation of $h(T)$ will be kept in our numerical computations; however, for the analytical estimations we will ignore it, simply assuming
\begin{equation}
	T(a)\propto \frac{1}{a}\,,
\end{equation}
which is valid up to variations in the number of relativistic degrees of freedom.
This scaling allows to find the frequencies $f_\text{re}$ and $f_{\star}$ corresponding, respectively, to the temperatures $T_\text{re}$ and $T_{\star}$ used in Eq.~\eqref{A(T)}:
\begin{eqnarray}
	f_\text{re}&=&\frac{k_\text{re}}{2\pi}=\frac{a_\text{re}\,H(a_\text{re})}{2\pi}=\frac{a_0}{2\pi}\frac{T_0}{T_\text{re}}H_\text{GR}(T_\text{re})=\frac{a_0}{3}\sqrt{\frac{\pi g}{5}}\,\frac{T_0\,T_\text{re}}{M_P}\,,\\
	f_{\star}&=&\frac{k_{\star}}{2\pi}=\frac{a_{\star}\,H(a_{\star})}{2\pi}=\frac{a_0}{2\pi}\frac{T_0}{T_{\star}}A(T_{\star})\,H_\text{GR}(T_{\star})\simeq\frac{a_0}{3}(1+\eta)\sqrt{\frac{\pi g}{5}}\frac{T_0\,T_{\star}}{M_P}\,,
\end{eqnarray}
for $T_{\star}\gg T_\text{re}$.
In the following, the two limiting cases corresponding to $T\ll T_\text{re}$ and $T\gg T_{\star}$ (or equivalently $f\ll f_\text{re}$ and $f\gg f_{\star}$) will be studied in detail.

\begin{figure}[t]
\begin{center}
\includegraphics[height=0.32\textwidth]{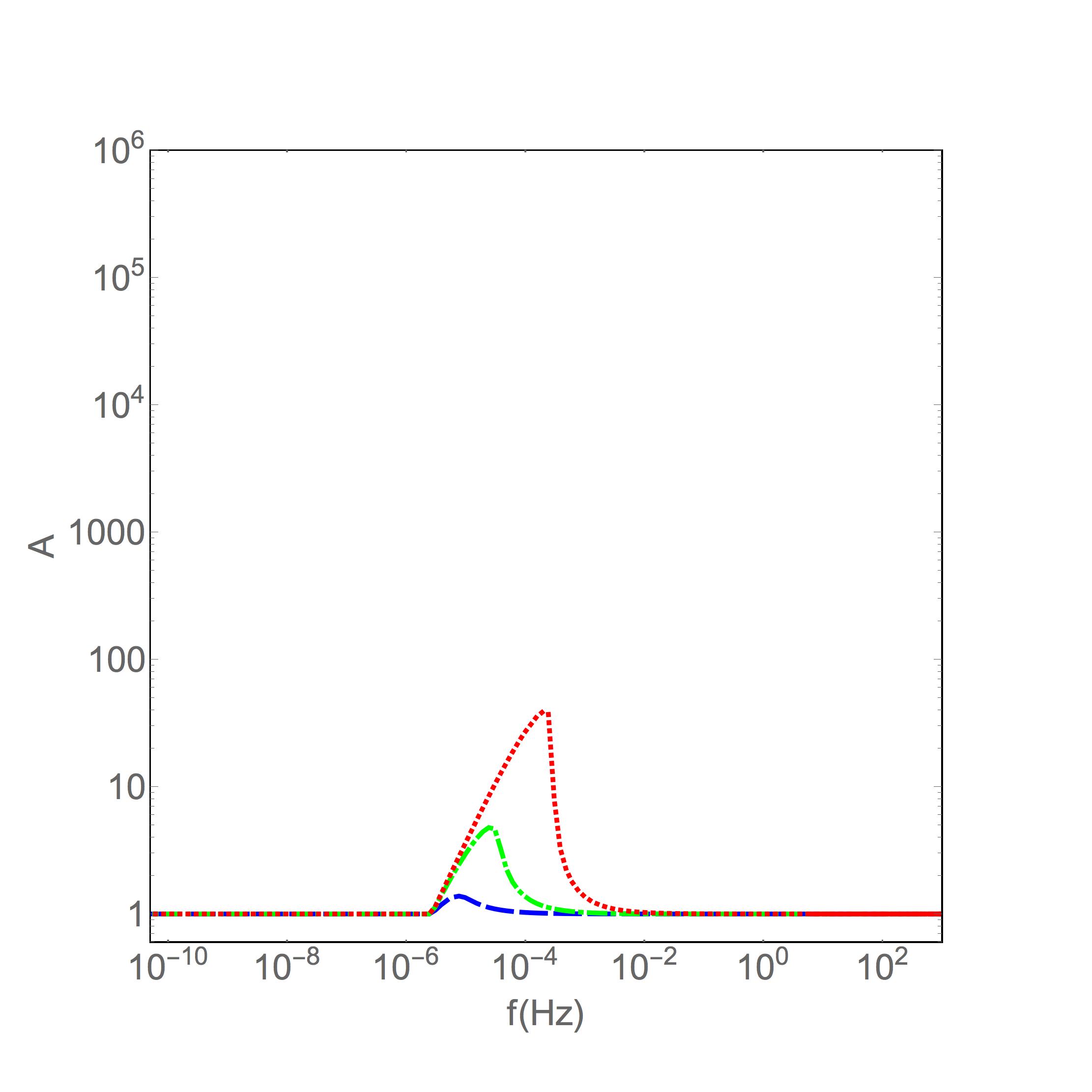}
\includegraphics[height=0.32\textwidth]{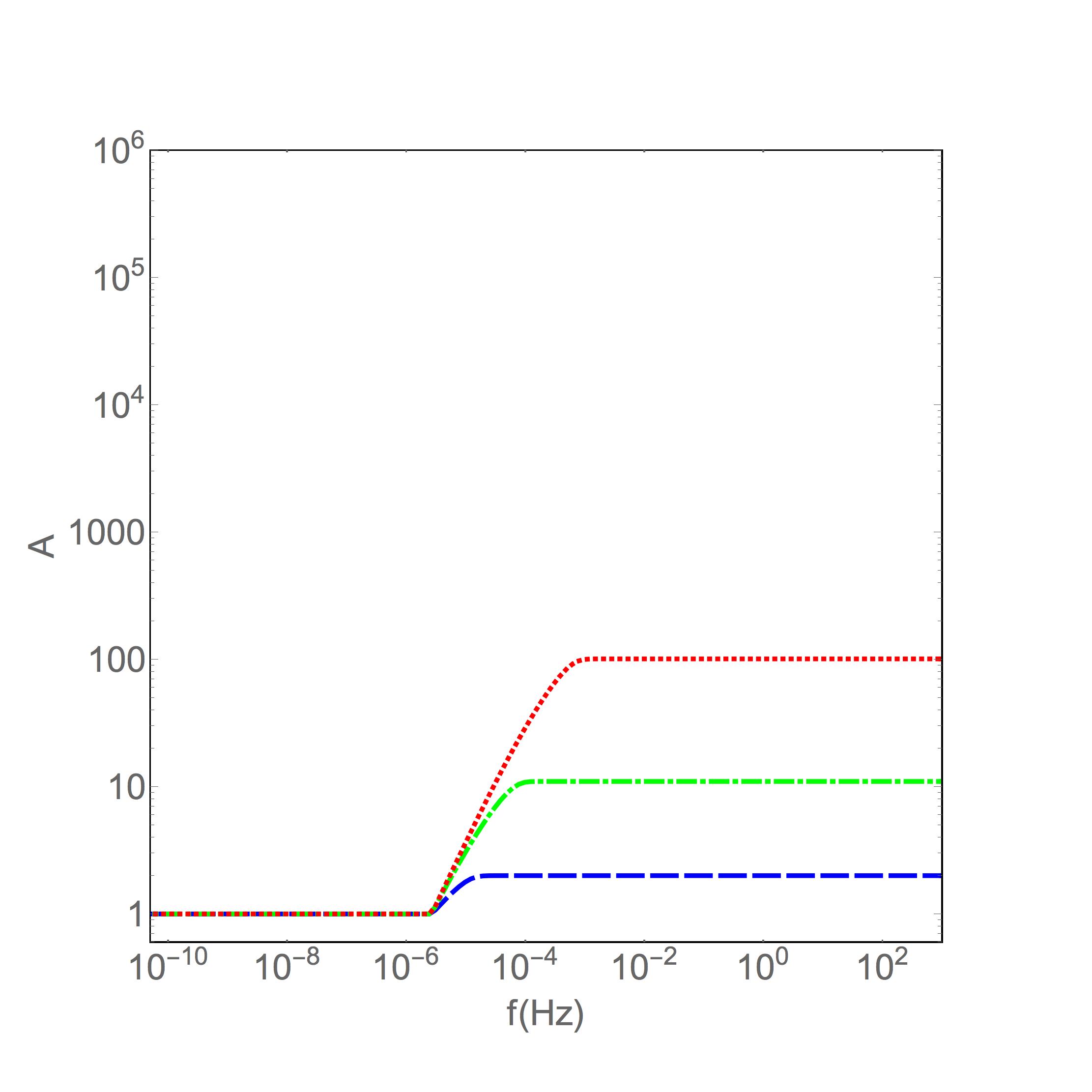}
\includegraphics[height=0.32\textwidth]{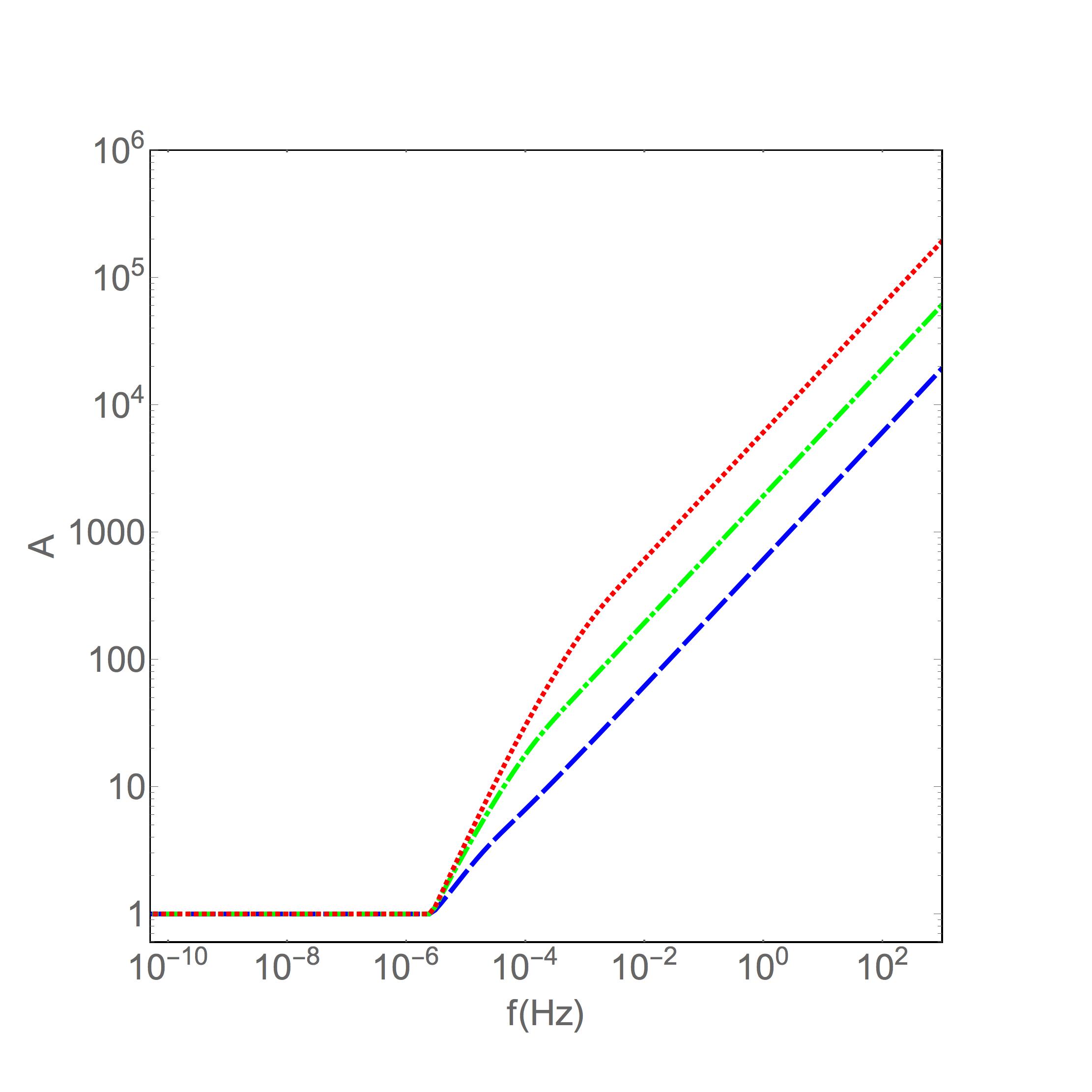}
\includegraphics[height=0.32\textwidth]{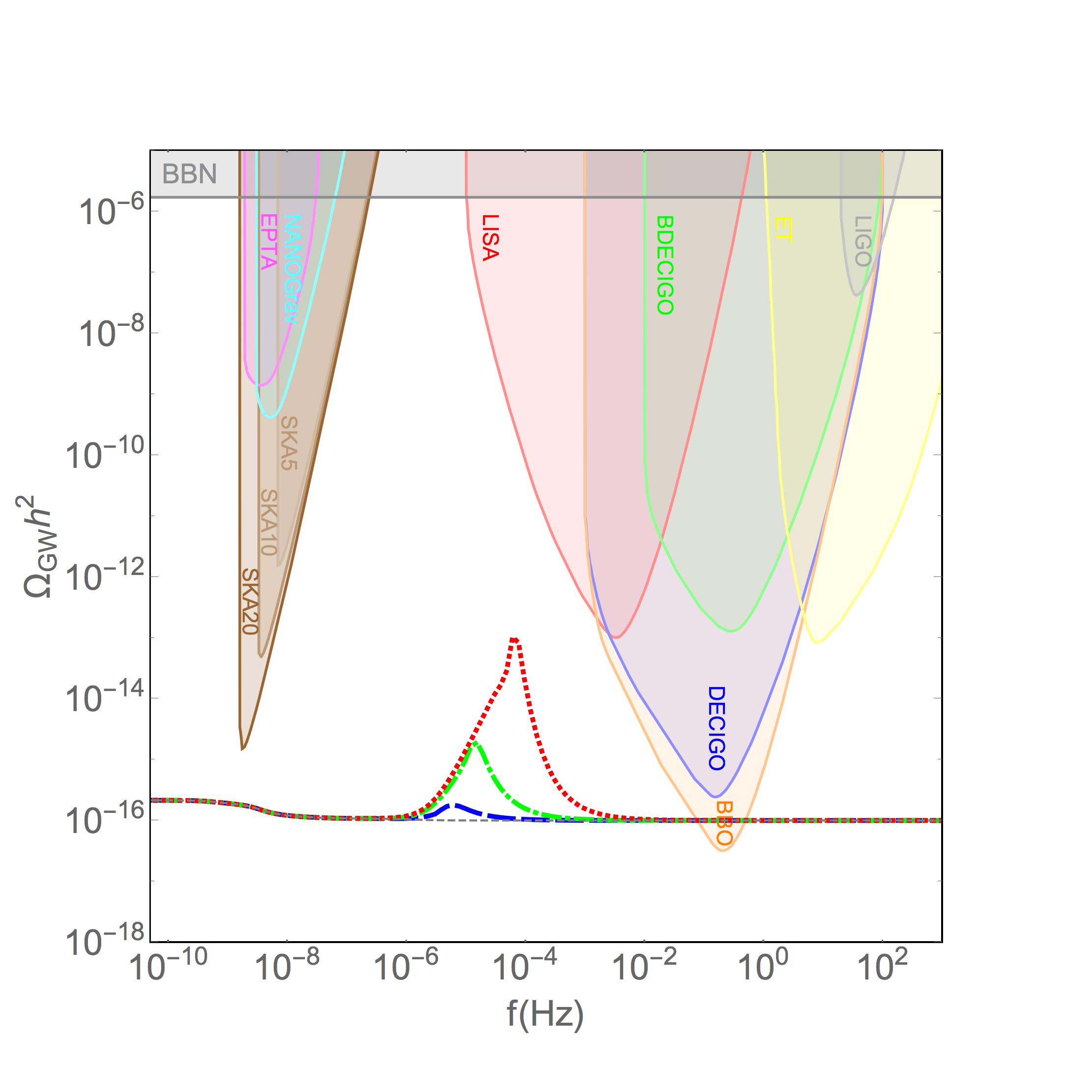}
\includegraphics[height=0.32\textwidth]{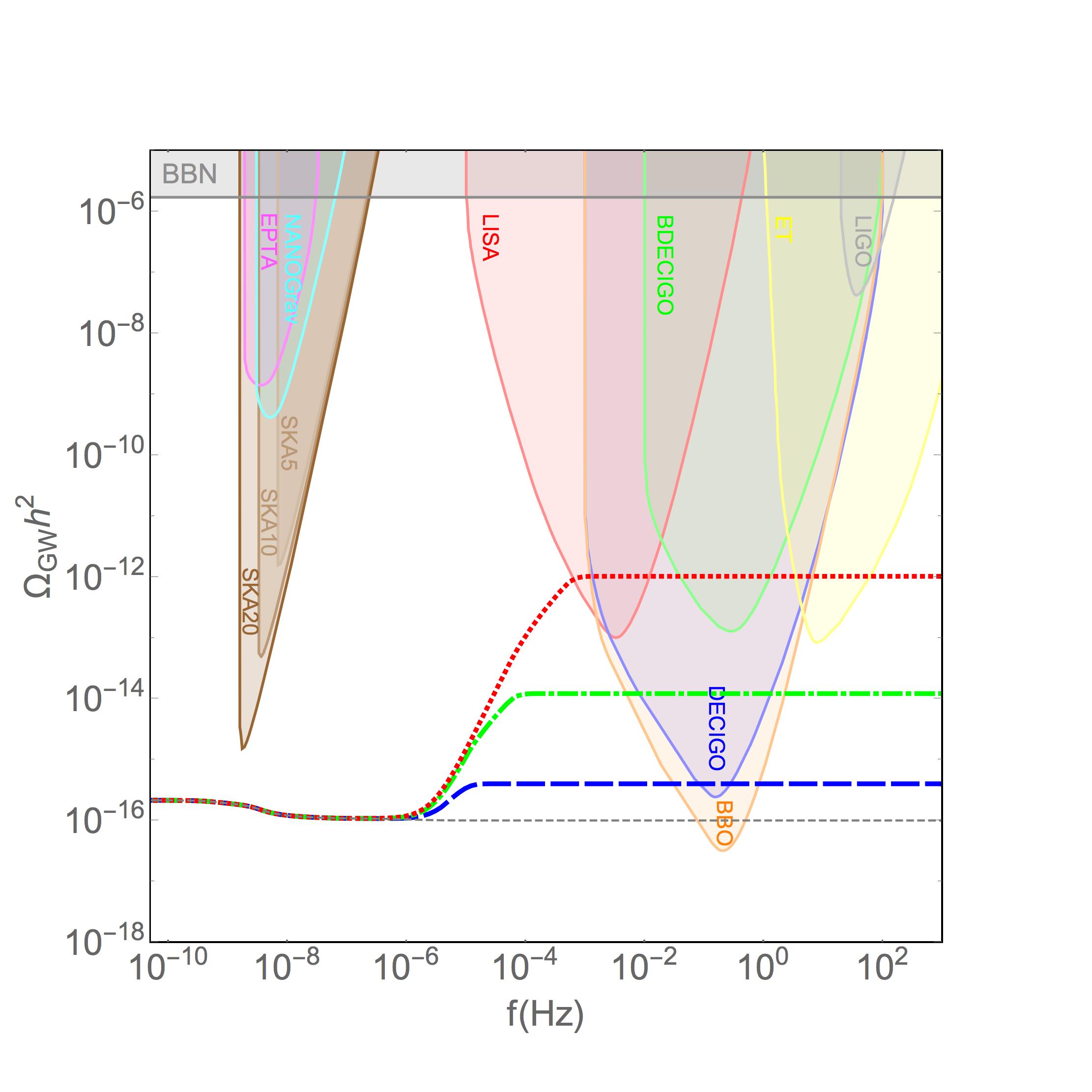}
\includegraphics[height=0.32\textwidth]{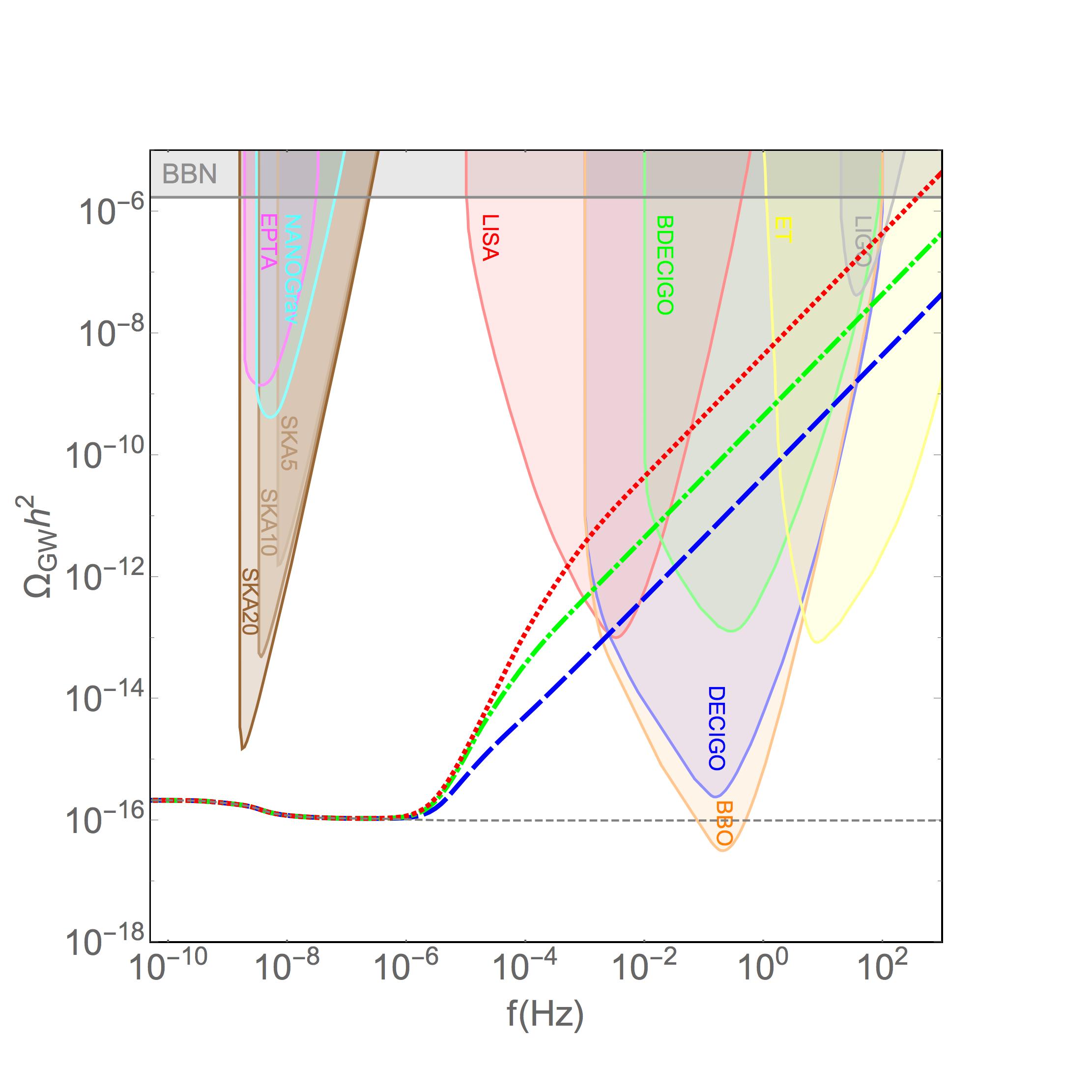}
	\caption{\it   Amplification factor $A$ (upper panels) and PGW spectrum $\Omega_\text{GW}h^2$ (lower panels) as a function of the frequency $f$ for $T_{\star}=T_\text{re}=100$~GeV (or equivalently $f_{\star}=f_\text{re}\simeq 2.5\times10^{-6}$~Hz) and $\eta=1$ (blue dashed lines), $\eta=10$ (green dot-dashed lines), $\eta=100$ (red dotted lines), and $\nu=-1$ (left panels), $\nu=0$ (central panels), $\nu=1$ (right panels).
	For reference, the PGW for standard cosmology is depicted with gray dashed lines, assuming a scale invariant primordial tensor spectrum ($n_T=0$) and a tensor-to-scalar ratio $r=0.07$.
	The colored regions in the lower panel correspond to projected sensitivities for various GW observatories, and to the BBN constraint described in the text.}\label{fig:omega}
\end{center}
\end{figure}

The upper panels of Fig.~\ref{fig:omega} show the amplification factor $A$ as a function of the frequency $f$ for $T_{\star}=T_\text{re}=100$~GeV (or equivalently $f_{\star}=f_\text{re}\simeq 2.5\times10^{-6}$~Hz), taking $\eta=1$ (blue dashed lines), $\eta=10$ (green dot-dashed lines), $\eta=100$ (red dotted lines), and $\nu=-1$ (left panels), $\nu=0$ (central panels), $\nu=1$ (right panels).

\subsection[$f\ll f_\text{re}$]{$\boldsymbol{f\ll f_\text{re}}$}
In the range of frequencies $f\ll f_\text{re}$, or equivalently for temperatures $T\ll T_\text{re}$, cosmology should converge to GR, and therefore before the onset of BBN one has that
\begin{equation}
	H(a)=H_\text{GR}(a)=H(a_\text{re})\left(\frac{a_\text{re}}{a}\right)^2\,,
\end{equation}
where $a_\text{re}$ is  the scale factor at $T=T_\text{re}$. 
The PGW relic density in Eq.~\eqref{pgw-relic} becomes
\begin{equation}
\label{omgw-gen-nu-posit}
	\Omega_\text{GW}(\tau_0,\,k)=\frac{\mathcal{P}_T(k)}{24}\left(\frac{a_{\text{re}}}{a_0}\right)^4\left(\frac{H(a_{\text{re}})}{H_0}\right)^2\propto \mathcal{P}_T(k)\,,
\end{equation}
showing the same  scale dependence as the primordial tensor power spectrum $\mathcal{P}_T(k)$, as expected from the standard cosmology.

\subsection[$f_\text{re} \ll f$]{$\boldsymbol{f_\text{re} \ll f}$}
Contrary to the previous case, in the range of frequency $f_\text{re}\ll f$ the amplification factor $A$ plays a major role.
In the following subsections, the regimes where $\nu>0$, $\nu=0$, and $\nu<0$ will be studied separately. 

\subsubsection*{Case $\boldsymbol{\nu>0}$}
If $\nu$ takes positive values, the Hubble rate can be expressed as
\begin{equation}
	H(a)\simeq H(a_\text{re})\left(\frac{a_\text{re}}{a}\right)^{2+\nu},
\end{equation}
which allows to express the PGW relic density in Eq.~\eqref{pgw-relic} as
\begin{equation}\label{hubnup}
	\Omega_\text{GW}(\tau_0,\,k)=\frac{\mathcal{P}_T(k)}{24\,a_0^4\,H_0^2}\left[H(a_\text{re})\,k^\nu\,a_\text{re}^{2+\nu}\right]^\frac{2}{1+\nu}
	\propto \mathcal{P}_T(k)\,k^\frac{2\nu}{1+\nu}.
\end{equation}
The PGW spectrum gains an extra factor $k^\frac{2\nu}{1+\nu}$, and is therefore blue-tilted with respect to the original tensor power spectrum.
This enhancement in the PGW spectrum can alternatively be understood by examining the friction term in Eq.~\eqref{eq:friction}:
\begin{equation} \label{dampfacnupos}
2\,\frac{a'}{a}=\frac{4}{1+3\omega}\,\frac{1}{\tau}\simeq\frac{2}{1+\nu}\frac{1}{\tau}\,.
\end{equation}
With respect to the standard case where the Universe is dominated by radiation ($\omega=1/3$), the friction term is reduced and therefore the PGW spectrum is enhanced for $\omega>1/3$ or equivalently $\nu>0$.

\subsubsection*{Case $\boldsymbol{\nu=0}$}
In this simple case, the Hubble rate is enhanced by a constant factor $A=1+\eta$.
The PGW spectrum is therefore not distorted, just showing an overall shift of $A^2$:
\begin{equation}\label{pgw-eta}
    \OGW(\tau_0,\,k)\simeq\frac{(1+\eta)^2}{24}\mathcal{P}_T(k)\left[\frac{a_\text{hc}}{a_0}\right]^4\left[\frac{H_\text{hc}}{H_0}\right]^2\propto \mathcal{P}_T(k)\,.
\end{equation}

\subsubsection*{Case $\boldsymbol{\nu<0}$}

If $\nu$ takes negatives values, both for low ($f\ll f_\text{re}$) and high frequencies ($f\gg f_\text{re}$) the amplification factor tends to 1.
However, it is interesting to note that $A$ reaches a maximum at $f=\bar f\gtrsim f_\text{re}$ given by
\begin{equation}
    A(\bar f)\simeq\eta\left(\frac{T_\text{re}}{T_{\star}}\right)^\nu.
\end{equation}
The PGW spectrum has the same tilt as the original tensor power spectrum, but featuring a characteristic bump at $k=\bar k=2\pi\,\bar f$, with an amplitude given by
\begin{equation}\label{negnuk}
    \OGW(\tau_0,\,\bar k)\simeq\frac{1}{24}\eta^2\left(\frac{T_\text{re}}{T_{\star}}\right)^{2\nu}\mathcal{P}_T(\bar k)\left[\frac{a_\text{hc}}{a_0}\right]^4\left[\frac{H_\text{hc}}{H_0}\right]^2.
\end{equation}

The lower panels of Fig.~\ref{fig:omega} show the PGW spectrum $\Omega_\text{GW}h^2$ as a function of the frequency $f$ for $T_{\star}=T_\text{re}=100$~GeV (or equivalently $f_{\star}=f_\text{re}\simeq 2.5\times10^{-6}$~Hz) and $\eta=1$ (blue dashed lines), $\eta=10$ (green dot-dashed lines), $\eta=100$ (red dotted lines), and $\nu=-1$ (left panels), $\nu=0$ (central panels), $\nu=1$ (right panels).
For reference, the PGW spectrum in the case of standard cosmology is depicted with gray dashed lines, assuming a scale-invariant primordial tensor spectrum ($n_T=0$) and a tensor-to-scalar ratio $r=0.07$.
The colored regions in the lower panel correspond to projected sensitivities for various GW observatories.
As expected from the analytical estimations in Eqs.~\eqref{hubnup}, \eqref{pgw-eta} and~\eqref{negnuk}, the PGW spectra are boosted due to the variation of the Hubble expansion rate.
Such boosts clearly follow the evolution of the amplification factor $A$.
The effect of the modified cosmologies could give a localized boost, an overall boost, or a change in the frequency dependence, for $\nu<0$, $\nu=0$, or $\nu>0$, respectively.

\section{PGW in scalar-tensor theories}
\label{sec:sttpgw}

Having described the impact of a general parameterization of modified gravity models on the PGW spectrum, we now concentrate on a simple model which allows for explicit calculations. We will like to understand analytically how the boost to the PGW spectrum is related to conformal parameters in a specific modified gravity model:
the ST theory of gravity.

ST theories are defined by the action $S\equiv S_\text{ST}+S_\text{m}$~\cite{Catena:2009tm} (see also Refs.~\cite{Modak:1999nm, Schelke:2006eg, Dent:2009bv, Leon:2013qh, Meehan:2015cna}),
 \begin{equation}\label{SSTT}
   S_\text{ST}=\frac{1}{16\pi\,G_*} \int d^4 x\sqrt{-{ g}} \left[F(\phi)\,{ R}(g)-Z(\phi)\,{g}^{\mu\nu}\,\partial_\mu \phi\,\partial_\nu\phi-2{ V}(\phi)\right],
 \end{equation}
where $R$ is the Ricci scalar, $F$ and $Z$ are arbitrary dimensionless functions of the field $\phi$ (also dimensionless),
and $S_\text{m}=S_\text{m}[\psi_\text{m},\,g_{\mu\nu}]$ is the matter action (here $\psi_m$ denotes the matter fields that couple to the metric tensor $g_{\mu\nu}$). The action~\eqref{SSTT} reduces to the well-known Brans-Dicke theory when $F(\phi)=\phi$, $Z(\phi)\propto \phi^{-1}$ and $V(\phi)=0$.
Additionally, let us note that this action is formulated in the Jordan frame.

The conformal transformation
\begin{equation} \label{conformal_transformation}
    g_{\mu\nu}=A_C^2(\phi_{*})\,g_{*\mu\nu}\,,
\end{equation}
together with the change of variables 
\begin{eqnarray}
    \left(\frac{d\phi_*}{d\phi}\right)^2&=&\frac34\left[\frac{d\ln F(\phi)}{d\phi}\right]^2+\frac{Z(\phi)}{2\,F(\phi)}\,,\\
    A_C(\phi_{*})&=&F^{-\frac12}(\phi)\,,\\
    V_{*}(\phi_{*})&=& \frac{V(\phi)}{2\,F^2(\phi)}\,,
\end{eqnarray}
yield the action in the Einstein frame
 \begin{equation}\label{SSTTE}
   S_\text{ST}=\frac{1}{16\pi G_*}\int d^4 x_{*} \sqrt{-g_{*}}\left[R_{*}(g_{*})-2g_{*}^{\mu\nu}\partial_\mu \phi_{*}\partial_\nu \phi_{*}-4V_{*}(\phi_{*})\right],
 \end{equation}
while $S_\text{m}=S_\text{m}[\psi_\text{m},\,A_C^2\,g_{*\mu\nu}]$.
The FLRW cosmological field equations in the Einstein frame  are given by~\cite{Catena:2004ba, Catena:2009tm, Meehan:2015cna}
\begin{eqnarray}\label{HUBSTT}
    &&3H_*^2\equiv 3\left(\frac{\dot a_{*} }{a_{*} }\right)^2 = 8\pi G_*\rho_{*} +{\dot \phi_{*}} ^2+2V_{*}(\phi_{*} )\,, \\
    &&3\frac{\ddot a_{*} }{a_{*}}= -4\pi G_*(\rho_{*} +3 p_{*}) -2{\dot \phi_{*}} ^2+2V_{*}(\phi_{*} )\,, \label{accSTT}\\
    &&{\ddot \phi_{*}} + 3H_{*}\,{\dot \phi_{*}}+\frac{dV_*}{d\phi_*} =-4 \pi G_*\, \alpha(\phi_*)\,\left(\rho_{*} -3p_{*} \right) \,, \label{phiSTT}
 \end{eqnarray}
where the dots denote derivatives with respect to the time variable $t_*$.
Deviations of ST theories from GR are parameterized by
\begin{equation}
    \alpha(\phi_{*})\equiv\frac{d\ln A_C(\phi_{*})}{d\phi_{*}}\,,
\end{equation}
where in the limit $\alpha\to 0$, $A_C$ becomes a constant, the two frames coincide, and therefore the ST theory reduces to GR.

We define the number of $e$-folds in the Einstein frame as  $N=\ln (a_*/a_{*0})$, where the subindex `0' labels quantities evaluated at present.
By definition, the conformal factor at present time is $A_C(\phi_{*0})=1$. 
Additionally, the relations between scale factor, time, energy density, and pressure in the Jordan and Einstein frames are
\begin{equation}\label{je-vars}
    a=A_C(\phi_{*})\,a_*\,,\quad
    dt=A_C(\phi_{*})\,dt_*\,,\quad
    \rho=\frac{\rho_*}{A_C^4(\phi_{*})}\,,\quad
    p=\frac{p_*}{A_C^4(\phi_{*})}\,.
\end{equation}
Therefore, the Hubble rates in the two frames are related by 
\begin{equation}\label{hub2}
    H=\frac{H_{*}+\alpha(\phi_{*})\,\dot{\phi}_{*}}{A_C(\phi_{*})}=H_*\frac{1+\alpha(\phi_{*})\,\frac{d\phi_*}{dN}}{A_C(\phi_{*})}\,,
\end{equation}
where the factor $1+\alpha(\phi_{*})\,\frac{d\phi_*}{dN}$ has to be positive.
Additionally, the relation between gravitational constants in GR and the Einstein frame is given by~\cite{Nordtvedt:1970uv, Damour:1996xx}
\begin{equation}\label{grconst}
    G=G_*\,A_C^2(\phi_{*0})\,\left[1+\alpha^2(\phi_{*0})\right].
\end{equation}
Finally, Eqs.~\eqref{phiSTT} and~\eqref{hub2} can be rewritten as%
\footnote{Hereafter we set $V_*=0$.}
\begin{eqnarray}\label{hubconf}
   H=\frac{A_C(\phi_{*})}{A_C(\phi_{*0})}\frac{1+\alpha(\phi_{*})\,\frac{d\phi_*}{dN}}{\sqrt{1+\alpha^2(\phi_{*0})}\,\sqrt{1-\frac13\left(\frac{d\phi_*}{dN}\right)^2}}H_\text{GR}\,, \\ 
    \frac{2}{3-\left(\frac{d\phi_*}{dN}\right)^2}\,\frac{d^2\phi_*}{dN^2}+ \left[1-\omega\right]\frac{d\phi_*}{dN}+\alpha(\phi_{*})\,[1-3\omega]=0\,,  \label{phiEq}
 \end{eqnarray}
where $\omega=\omega(T)$ varies from $1/3$ to $-1$ after reheating.
In particular, during radiation-domination era, its evolution is given by the variation of the effective number of degrees of freedom
\begin{equation}
   \omega(T)=\frac43\frac{h(T)}{g(T)}-1\,.
\end{equation}
Let us note that $\omega$ has to be understood as the equation-of-state parameter of the SM bath.
That means that after neutrino decoupling, only photons and electrons/positrons contribute to $\omega$.
\begin{figure}[t]
    \begin{center}
        \includegraphics[height=0.4\textwidth]{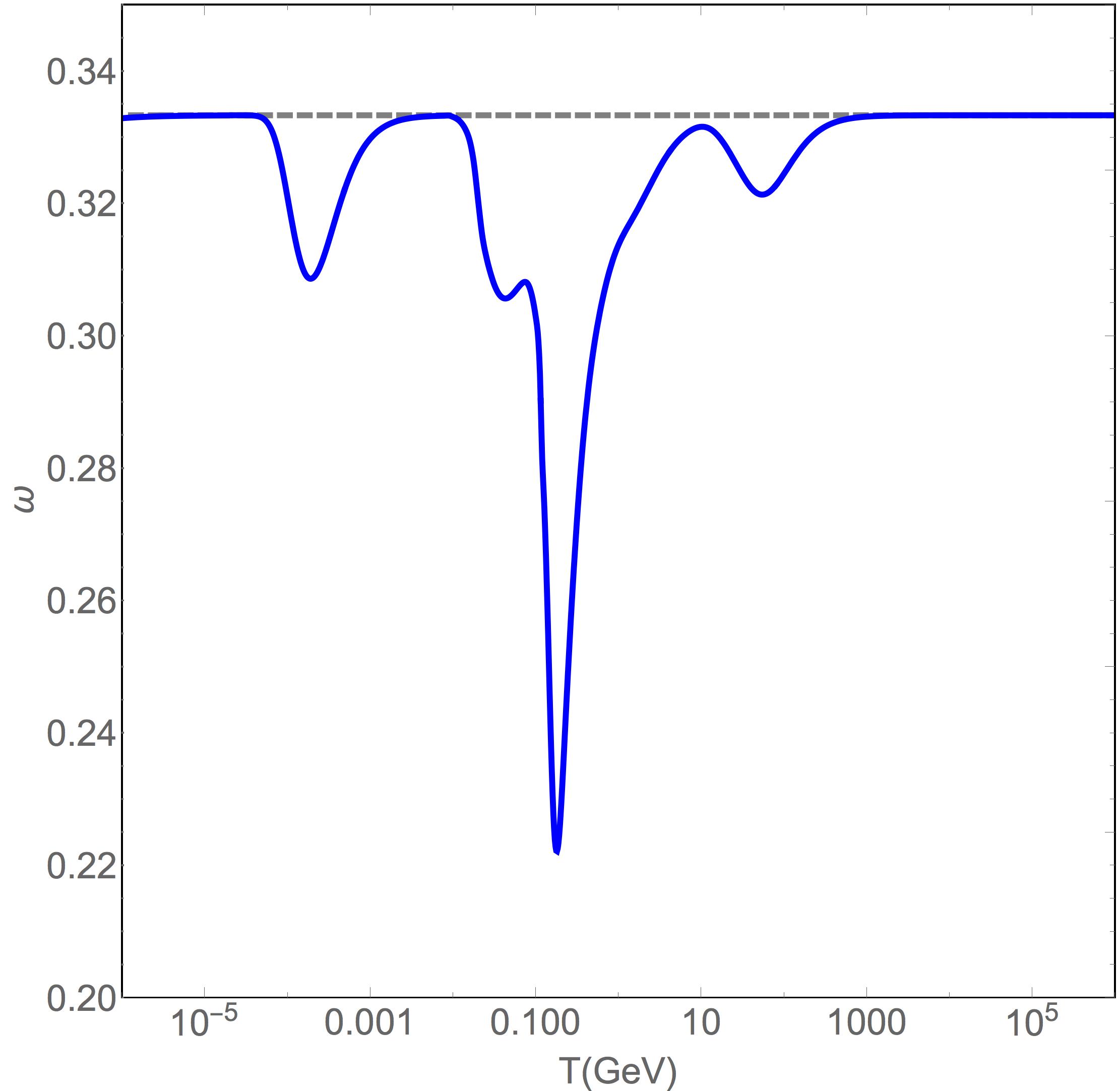}
	    \caption{\it  Evolution of the equation-of-state parameter $\omega$ with respect to the photon temperature $T$.
	    For reference, $\omega=1/3$ corresponding to radiation domination is also shown.}
        \label{fig:EoS}
    \end{center}
\end{figure}
Figure~\ref{fig:EoS} shows the evolution of the equation of state as a function of $T$.
Deviations from $\omega=1/3$ correspond to thresholds where SM particles become non-relativistic and to the effect of particle interactions.
In particular, the funnels at $T\simeq 0.5$~MeV, $T\simeq 150$~MeV, and $T\simeq 100$~GeV correspond to the neutrino decoupling, QCD crossover, and electroweak crossover, respectively \cite{Drees:2015exa}.
Finally, by noticing that entropy is conserved in the Jordan frame, the evolution of the SM temperature is given by
\begin{equation}\label{tempsttapp}
    T= \left[\frac{h(T_0)}{h(T)}\right]^{\frac13}\frac{A_C(\phi_{*0})}{A_C(\phi_{*})}\,T_0\, e^{-N}\,.
\end{equation}
 
Going further in the analysis requires fixing the conformal factor.
One usual choice corresponds to~\cite{Damour:1992we, Damour:1993hw, Meehan:2015cna,Coc:2006rt}
\begin{equation}\label{ACdef}
    A_C(\phi_{*})=e^{\frac12 \beta\,\phi_{*}^2},
\end{equation}
which implies that $\alpha(\phi_{*})=\beta\,\phi_{*}$.
In our numerical study, the ST model is fully set by fixing $\beta$, the initial value of the field $\phi_{*\text{in}}$ and its derivative $\left(d\phi/dN\right)_{*\text{in}}$, at a high temperature $T_\text{in}=10^{14}$~GeV.
For the sake of simplicity, here we focus on the case $\left(d\phi/dN\right)_{*\text{in}}=0$.
Additionally, the specific choice of $T_\text{in}$ is not important, as long as it is much higher than the electroweak scale.
In fact, as it will be seen, for $T\gg T_\text{EW}$ the field $\phi_*$ does not evolve.

To analytically understand the behavior of the Hubble expansion rate and the PGW spectrum, it is convenient to introduce two scales, one being the electroweak and the other the BBN scale.
The corresponding frequencies are
\begin{eqnarray}
    f_\text{EW} &=&\frac{k_\text{EW}}{2\pi}\simeq \frac{a_0}{3}\sqrt{\frac{\pi g(T_{\text{EW}})\,G}{5}}\,T_0\, T_{\text{EW}}\,e^{\frac12 \beta\phi_{\text{in}}^2} \,,\\
    f_\text{BBN} &=&\frac{k_\text{BBN}}{2\pi}\simeq\frac{a_0}{3}\sqrt{\frac{\pi g(T_\text{BBN})\,G}{5}}\,T_0\,T_\text{BBN}\,.
\end{eqnarray}
With respect to the two scales previously introduced, we study the following cases:

\subsubsection*{Case  $\boldsymbol{f\leq f_\text{BBN}}$}
For small frequencies $f\leq f_\text{BBN}$, or equivalently low temperatures $T\leq T_\text{BBN}$ (but still higher than the matter-radiation equality), the Universe energy density is dominated by SM radiation and therefore
\begin{equation}\label{stgwlowrelic}
    \Omega_\text{GW}(\tau_0,\,k)\simeq \frac{\mathcal{P}_T(k)}{24}\left(\frac{a_\text{BBN}}{a_0}\right)^4\left(\frac{H(a_\text{BBN})}{H_0}\right)^2\propto \mathcal{P}_T(k) \,,
\end{equation}
meaning that the PGW spectrum keeps the same scale dependence as the primordial tensor power spectrum $\mathcal{P}_T(k)$, as expected from the standard cosmology.

\subsubsection*{Case $\boldsymbol{f_{\text{EW}}\ll f}$}

In the opposite limit, for temperatures higher than $T_\text{EW}$ we are deep in the radiation-dominated era, with an equation of state constant and equal to $1/3$.
Therefore, the field $\phi_*$ is not rolling, staying at the value $\phi_*=\phi_{*\text{in}}$.
The Hubble expansion rate reduces to
\begin{equation}\label{sttpgwewhub}
    H \simeq e^{\frac12 \beta\,\phi_{*\text{in}}^2}\,H_\text{GR}\,, 
\end{equation}
where $\phi_{*0}=0$ was taken to recover GR at late times.
The spectrum of PGW is given by
\begin{equation}\label{sttpgwewspec}
    \Omega_\text{GW}(\tau_0,\,k)\simeq \frac{\mathcal{P}_T(k)}{24}\left(\frac{a_{\text{EW}}}{a_0}\right)^4\left(\frac{H(a_{\text{EW}})}{H_0}\right)^2e^{\beta\,\phi_{*\text{in}}^2}
    \propto e^{\beta\,\phi_{*\text{in}}^2}\,\mathcal{P}_T(k) \,,
\end{equation}
where $a_\text{EW}$ is the scale factor at $T=T_\text{EW}$.
Equation~\eqref{sttpgwewspec} shows an overall boost factor of $e^{\beta\,\phi_{*\text{in}}^2}$ on the PGW spectrum, and hence the same scale dependence as the primordial tensor power spectrum.

\subsubsection*{Case $\boldsymbol{f_\text{BBN}< f< f_\text{EW}}$}

Once the Universe cools down to temperatures $T\simeq T_\text{EW}$, SM particles start to become non-relativistic.
In fact, when the temperature drops below the mass of each of the particle types, the equation of state decreases, significantly differing from $1/3$ (see Fig.~\ref{fig:EoS}), and therefore an important reduction of $\phi_*$ takes place induced by the term proportional to $\alpha\,(1-3\omega)$ in Eq.~\eqref{phiEq}.

\begin{figure}[t]
    \begin{center}
        \includegraphics[height=0.42\textwidth]{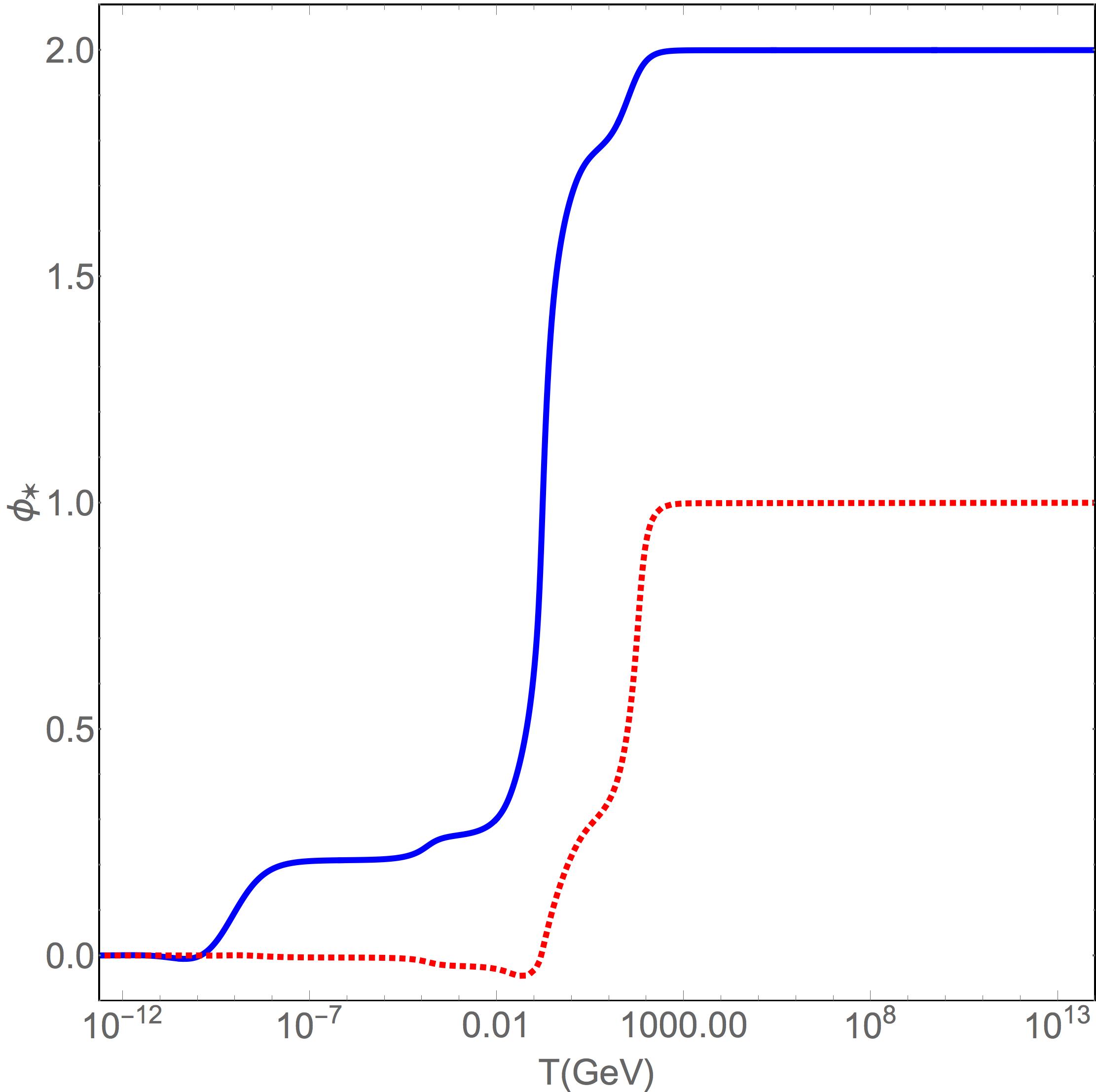}
        \includegraphics[height=0.42\textwidth]{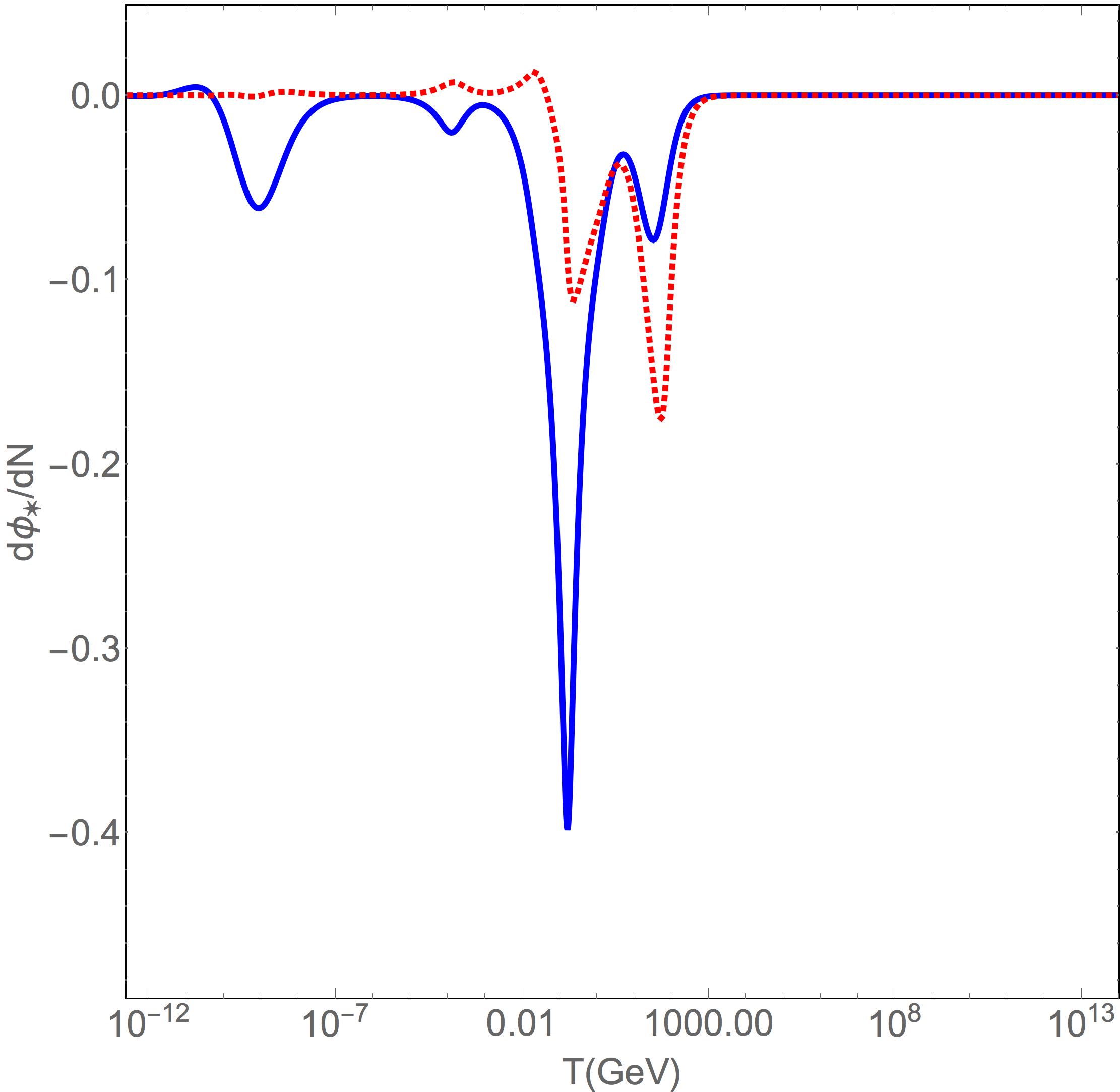}
        \caption{\it  Evolution of $\phi_*$ (left panel) and $d\phi_*/dN$ (right panel) with respect to the frequency $f$, for the benchmark points $[\beta,\,\phi_{*\text{in}}] = [1,\,2]$ (blue solid line) and $[5,\,1]$ (red dotted line).
        We also took $T_\text{in}=10^{14}$~GeV and $(d\phi_{*}/dN)_\text{in}=0$. 
        }
        \label{fig:scaltenphit}
    \end{center}
\end{figure}

Figure~\ref{fig:scaltenphit} shows the nontrivial evolution of the scalar field $\phi_{*}$ (left panel) and its derivative $d\phi_{*}/dN$ (right panel) as a function of the temperature $T$, for the benchmark points $[\beta,\,\phi_{*\text{in}}] = [1,\,2]$ (blue solid line) and $[5,\,1]$ (red dotted lines).
The figure shows that at high temperatures $T>T_\text{EW}$, the field stays at its initial value $\phi_{*\text{in}}$.
At the electroweak crossover, the reduction of the equation-of-state parameter induces a relaxation of the field that starts to roll to the minimum of its potential. 
We notice that its velocity $d\phi_*/dN$ tends to track the evolution of $\omega$.
Finally, at low temperatures $T\lesssim T_\text{BBN}$ the field $\phi_*\to 0$ and therefore GR is recovered \cite{Coc:2006rt,Meehan:2015cna}. 
In this way, the constraint on the speed of GW from LIGO is naturally avoided.

\begin{figure}[t]
    \begin{center}
        \includegraphics[height=0.5\textwidth]{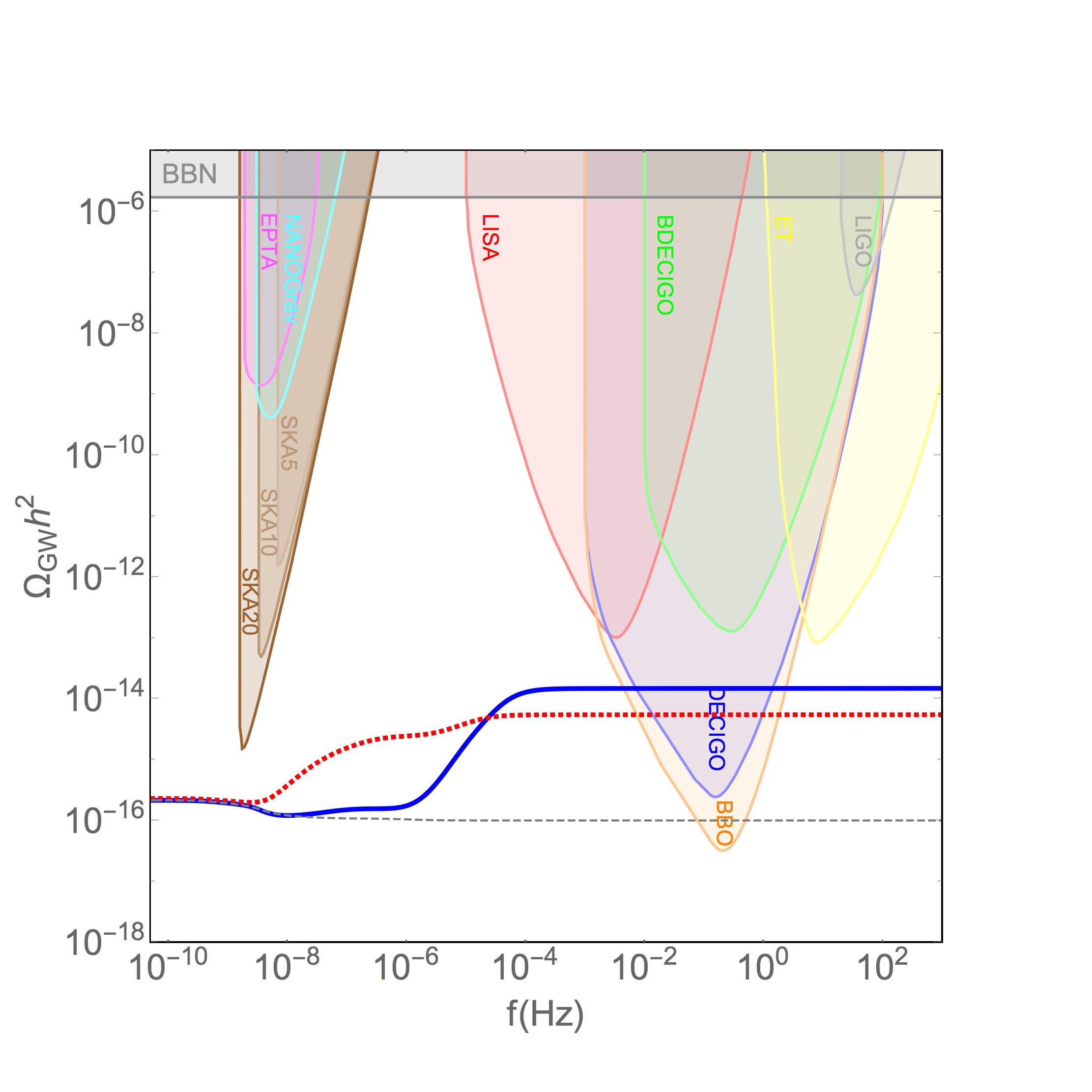}
        \caption{\it Scalar-Tensor Gravity: PGW spectrum, for the benchmark points $[\beta,\,\phi_{*\text{in}}] = [1,\,2]$ (blue solid line) and $[5,\,1]$ (red dotted line), and assuming a scale invariant primordial tensor spectrum ($n_T = 0$) with a tensor-to-scalar ratio $r = 0.07$.
        We also took $T_\text{in}=10^{14}$~GeV and $(d\phi_{*}/dN)_\text{in}=0$.
        For reference, the gray dashed line shows the PGW in the case of GR.
        The colored regions correspond to projected sensitivities for various GW observatories, and to the BBN constraint described in the text.}
        \label{fig:scaltengw}
    \end{center}
\end{figure}

Figure~\ref{fig:scaltengw} shows the corresponding PGW spectra produced for the same benchmark points used in Fig.~\ref{fig:scaltenphit} (i.e. $[\beta,\,\phi_{*\text{in}}] = [1,\,2]$ (blue solid line) and $[5,\,1]$ (red dotted line)), and assuming a scale invariant primordial tensor spectrum ($n_T = 0$) with a tensor-to-scalar ratio $r = 0.07$.
For reference, the gray dashed line shows the PGW in the case of GR.
Additionally, the colored regions correspond to projected sensitivities for various GW observatories, and to the BBN constraint described in the text.
For frequencies $f\lesssim f_\text{BBN}$, the PGW spectrum follows the one in GR. However, for $f> f_\text{BBN}$ there is a sizable scale-dependent boost that becomes constant at $f\simeq f_\text{EW}$.


\section{PGW in braneworld cosmology}
\label{branegw}

Here we investigate braneworld cosmological models as another example of modified gravity theories, to understand its impact on the spectrum of PGW.%
\footnote{In these models, other cosmological aspects as the impact in the DM relic density have been intensively investigated in Refs.~\cite{Okada:2004nc, Nihei:2004xv, Nihei:2005qx, AbouElDahab:2006glf, Meehan:2014bya, Baules:2019zwk}.}
In the braneworld cosmology, the Friedmann equation for a spatially flat Universe is found to be~\cite{Binetruy:1999ut, Shiromizu:1999wj, Binetruy:1999hy, Ida:1999ui}
\begin{equation}\label{hubbrane}
    H^2=\frac{8\pi G}{3}\rho\left(1+\frac{\rho}{\sigma}\right),
\end{equation}
where $\rho$ is the SM energy density and the parameter $\sigma$ is the brane tension.%
\footnote{For simplicity, we set the 4-dimensional cosmological constant and the so-called dark radiation parameter to zero~\cite{Ichiki:2002eh}.}
Then the brane tension parameter is related to the 5-dimensional Planck mass $M_5$ as 
\begin{equation}\label{brtension}
    \sigma\equiv 96\pi\,G\,M_5^6\,.
\end{equation}

To have a Universe dominated by SM radiation at $T=T_\text{BBN}$, it is required that $\sigma\gg\rho(T_\text{BBN})$.
It is therefore possible to define an effective temperature scale $T_\sigma$ from which the contribution of brane tension becomes important 
\begin{equation}\label{tempsigma}
    \sigma= \frac{\pi^2}{30}\,g(T_{\sigma})\,T_{\sigma}^4\,. 
\end{equation}
The Hubble expansion rate can be approximated as
 \begin{equation}\label{hubbr1}
    H^2\simeq\begin{cases}
    \frac{8\pi G}{3}\rho & \qquad\text{ for }  T\ll T_{\sigma} \,,\\[8pt]
    \frac{8\pi G}{3} \frac{\rho^2}{\sigma} & \qquad \text{ for } T\gg T_{\sigma} \,.
    \end{cases}
 \end{equation}
Taking into account that in braneworld cosmologies the entropy is conserved, it is possible to find the frequency $f_\sigma$ corresponding to the temperature $T_\sigma$:
\begin{equation}
\label{eq:fsigma}
    f_\sigma=\frac{k_\sigma}{2\pi}
    =\frac{a_0}{\pi}\frac{T_0}{T_\sigma}H_\text{GR}(T_\sigma)=\frac23\sqrt{\frac{\pi\,g}{5}}\frac{T_0\,T_\sigma}{M_P}.
\end{equation}

The PGW spectrum in the limiting cases presented in Eq.~\eqref{hubbr1} become:
 
 \subsection*{Case $\boldsymbol{f\ll f_{\sigma}}$}
At small frequencies $f\ll f_{\sigma}$, or equivalently low temperatures $T\ll T_{\sigma}\ll T_\text{BBN}$, the Universe energy density is dominated by SM radiation and therefore
 \begin{equation} \label{omradfsig}
 \Omega_\text{GW}(\tau_0,\,k)\propto \mathcal{P}_T(k) \,,
 \end{equation}
 meaning that the PGW spectrum keeps the same scale dependence as the primordial tensor power spectrum $\mathcal{P}_T(k)$, as expected from standard cosmology.
 
 \subsection*{Case $\boldsymbol{f\gg f_{\sigma}}$}
However, for high frequencies $f\gg f_{\sigma}$, Eq.~\eqref{pgw-relic} can be rewritten as
\begin{equation} \label{ommodfsig}
	\Omega_\text{GW}(\tau_0,\,k)=\frac{P_T(k)}{24\,a_0^4\,H_0^2}\left[H(a_\sigma)\,k^2\,a_\sigma^4\right]^\frac23
	\propto \mathcal{P}_T(k)\,k^\frac43.
\end{equation}
The PGW spectrum gains an extra factor $k^\frac43$, and is therefore blue-tilted with respect to the original tensor power spectrum, as expected from Eq.~\eqref{hubnup} in the case $\nu=2$.

\begin{figure}
\begin{center}
\includegraphics[height=0.49\textwidth]{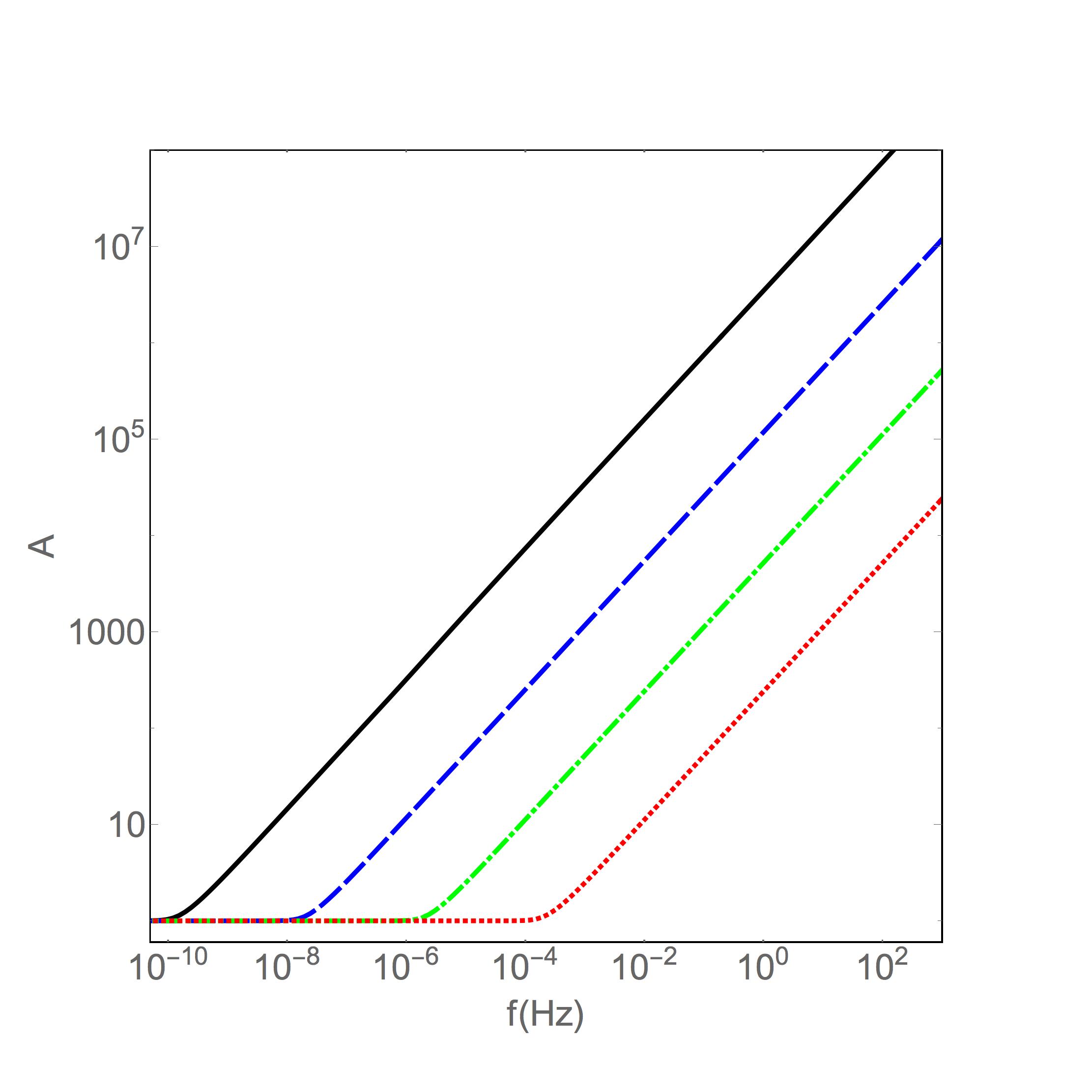}
\includegraphics[height=0.49\textwidth]{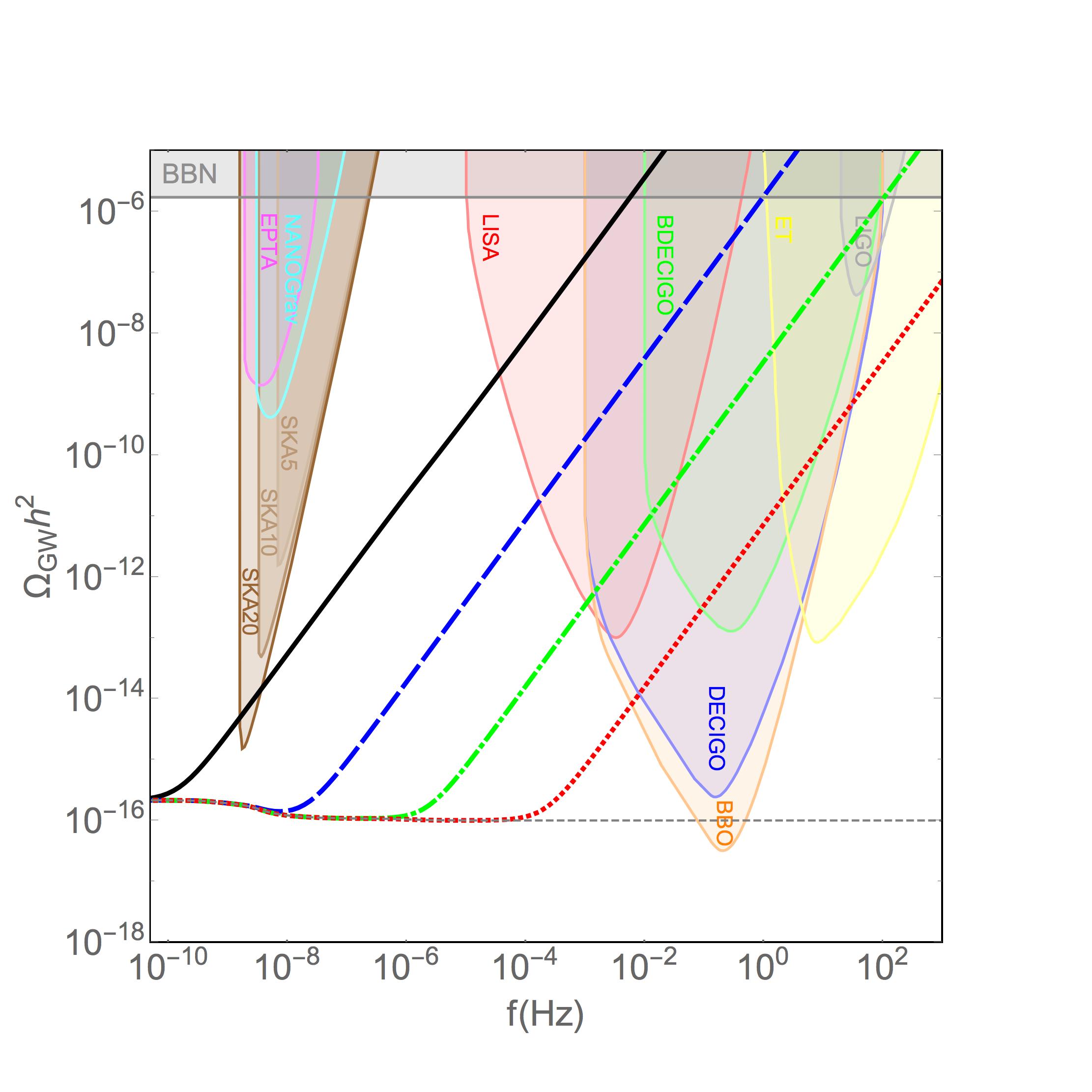}
	\caption{\it   Braneworld Gravity: Amplification factor $A$ (plot above) and PGW spectrum $\Omega_\text{GW}h^2$ (plot below) as a function of the frequency $f$ for $T_\sigma=10$~MeV (black  solid  lines), $T_\sigma=1$~GeV (blue dashed lines),  $T_\sigma=100$~GeV (dot dashed green lines) and $T_\sigma=10$~TeV (dotted red lines).
	For reference, the PGW for the standard cosmology case is depicted with gray dashed lines, assuming a scale invariant primordial tensor spectrum ($n_T = 0$) and a tensor-to-scalar ratio $r = 0.07$.
	The colored regions in the lower panel correspond to projected sensitivities for various GW observatories, and to the BBN constraint described in the text.
}
\label{fig:gwbc}
\end{center}
\end{figure}

Figure~\ref{fig:gwbc} shows the amplification factor $A$ (left panel) and the PGW spectrum $\Omega_\text{GW}h^2$ (right panel) as a
function of the frequency $f$ for $T_\sigma=10$~MeV (black  solid  lines), $T_\sigma=1$~GeV (blue dashed lines),  $T_\sigma=100$~GeV (dot dashed green lines) and $T_\sigma=10$~TeV (dotted red lines).
For reference, the PGW for standard cosmology is depicted with gray dashed lines, assuming a scale-invariant primordial tensor spectrum ($n_T = 0$) and a tensor-to-scalar ratio $r = 0.07$.
The colored regions in the lower panel correspond to projected sensitivities for various GW observatories, and to the BBN constraint described in the text.

 LIGO observations at late time posit an interesting bound on the parameter $T_{\sigma}$, that can be derived from Eqs.~\eqref{eq:fsigma} and~\eqref{ommodfsig}. In fact, 
\begin{equation}
    \Omega_{\text{GW, LIGO}}h^2 (f\sim10~\text{Hz}) \simeq 2\times 10^{-8} \simeq 2\times10^{-16} \left(\frac{r}{0.07}\right)\left(\frac{f}{f_{\sigma,\,\text{min}}}\right)^{4/3}.
\end{equation}
This gives $T_{\sigma,\,\text{min}}\simeq 300$~GeV and $f_{\sigma,\,\text{min}}\simeq 10^{-5}$~Hz as the minimum values of characteristic temperature and frequency allowed by LIGO.
 
\section{Summary and conclusions}
\label{sec:conclusion}

Modified cosmologies, based on UV-completions of GR, predict modified expansion histories in the early Universe, which are typically unconstrained by cosmological observations hitherto.  However, current and upcoming GW detectors will open up the possibility of directly probing such era via PGW.

We investigated modified gravity theories which generate nonstandard Hubble expansions of the Universe in the pre-BBN epoch, looking for possible enhancements in the relic density of the PGW spectrum. %
In Sec.~\ref{sec:modgragen} we considered a general parametrization for various theories of gravity affecting the expansion of the Universe.
Examples of modified PGW spectra were shown in Fig.~\ref{fig:omega}.
In Sec.~\ref{sec:sttpgw} we focused on the case of scalar-tensor theories that modify Einstein's gravity due to an additional scalar degree of freedom which couples to the Ricci scalar.
Fig.~\ref{fig:scaltengw} shows examples of typical PGW spectra enhanced by scalar-tensor cosmologies, which can be detectable by DECIGO.
The same behavior happens in braneworld scenarios described in Sec.~\ref{branegw}.
In fact, depending on the values of brane tension parameter $\sigma$, one may get an enhanced spectrum of the PGW, as shown in Fig.~\ref{fig:gwbc}.

Finally, GW experiments provide complementary tests of gravity in the early Universe and enlighten us what modifications to Einstein gravity may really explain phenomena over different cosmic scales consistently. In addition to other gravity tests, PGW detection will narrow down the parameter space for the modification of GR in the early Universe. However, one requires to do a detailed analysis of how the bounds from such laboratory/astrophysical probes of gravity compete/complement that by PGW, any such analysis is beyond the scope of our current paper.

 The analysis presented in this paper should only be regarded as a first step in the quest of modified gravity probes in the pre-BBN era via PGW. Such modifications in the early universe lead to modifications on the dark matter relic density, the matter-antimatter asymmetry, and the abundance of primordial black holes and microhalos.

Overall, the cosmological era between the end of inflation and the beginning of radiation-dominated era is unknown and remains largely unexplored. Several UV-complete scenarios motivate nonstandard cosmology or modified gravity epochs prior to BBN. The present investigation is timely because several decades of frequency ranges of various GW amplitudes should be accessible to the next generation GW experiments. Consequently, additional constraints or signals from these experiments could point to new physics in the era prior to BBN. 

For future investigations in this direction, we would like to relax our assumptions regarding the initial primordial tensor spectrum generated during inflation, and correlate the PGW probes with other tests-of-gravity experiments.%
\footnote{See Ref.~\cite{Baker:2014zba} for such complementary probes of gravity between test-of-gravity experiments and GW of astrophysical origin.} We believe that with the network of GW detectors that are planned for heralding and advancing GW astronomy, it will soon be possible to make the dream of understanding UV completion of gravity a reality.  

\acknowledgments
We acknowledge Arindam Chatterjee, Debatri Chattopadhyay, Mohammad Ali Gorji,  Anupam Mazumdar, and J{\"u}rgen Schaffner-Bielich for useful discussions.
NB is partially supported by Spanish MINECO under Grant FPA2017-84543-P, and from Universidad Antonio Nariño grants 2018204, 2019101, and 2019248. 
FH is supported by the Deutsche Forschungsgemeinschaft (DFG, German Research Foundation) - Project number 315477589 - TRR 211.

\bibliography{main}
\end{document}